\documentclass{elsarticle}

\usepackage{amsmath}
\usepackage{bm}
\usepackage{tikz}
\usepackage{pgfplots}
\usepgflibrary{arrows}
\usepackage{rotating} 
\graphicspath{{pictures/}}

\newproof{pf}{Proof}

\journal{arXiv.org}

\begin{document}

\begin{frontmatter}

\title{Numerical simulation of oxidation processes in a cross--flow around tube bundles}
 
\author[nsi]{Alexander~G.~Churbanov}
\ead{achur@ibrae.ac.ru}

\author[itwm]{Oleg~Iliev}
\ead{iliev@itwm.fraunhofer.de}

\author[nsi]{Valery~F.~Strizhov}
\ead{vfs@ibrae.ac.ru}

\author[nsi,svfu]{Petr~N.~Vabishchevich\corref{cor}}
\ead{vabishchevich@gmail.com}

\address[itwm]{Fraunhofer Institute for Industrial Mathematics ITWM,  Fraunhofer-Platz 1, D-67663 Kaiserslautern, Germany}

\address[nsi]{Nuclear Safety Institute, Russian Academy of Sciences, 52, B.~Tulskaya, Moscow, Russia}

\address[svfu]{North--Eastern Federal University, 58, Belinskogo, Yakutsk, Russia}

\cortext[cor]{Corresponding author}

\begin{abstract}
An oxidation process is simulated for a bundle of metal tubes in a cross--flow.
A fluid flow is governed by the incompressible Navier--Stokes equations.
To describe the transport of oxygen, the corresponding convection--diffusion equation is applied.
The key point of the model is related to the description of oxidation processes taking into account the growth 
of a thin oxide film in the quasi--stationary approximation.
Mathematical modeling of oxidant transport in a tube bundle is carried out in the 2D approximation. 
The numerical algorithm employed in the work is based on the finite--element discretization in space 
and the fully implicit discretization in time.
The tube rows of a bundle can be either in--line or staggered in the direction of the fluid flow velocity. 
The growth of the oxide film on tube walls is predicted for various bundle structures 
using the developed oxidation model.
\end{abstract}


\end{frontmatter}

\section{Introduction}

Many industrial applications involve fluid flows through tube structures.
An example is the heat exchange equipment for nuclear power plants.
A study of heat and mass transfer in these systems is of great practical interest
\citep{kays1998compact,bergman2011fundamentals}. 
In addition to large--scale experimental studies \citep{vzukauskas1972heat,iwaki2004piv,paul2007experimental},
computational technologies are in common use for theoretical considerations
of key aspects of heat and mass transfer processes in tube bundles
\citep{beale1999numerical,wang2000analysis,el2005analysis,jayavel2009numerical}.

In numerical simulation of cross--flows around a bundle of tubes, the emphasis is
on predicting hydrodynamic and thermal phenomena. Using finite--difference, finite--volume or finite--element methods, 
laminar or turbulent flows are studied at various technological conditions (see, for example,
\cite{gowda1998finite,zdravistch1995numerical,dhaubhadel1987finite,wilson2000modeling,khan2006convection,lam2010experimental,benarji2008unsteady,horvat2006heat,wang2010turbulence,tahseen2015overview} 
and the bibliography cited in these works).

To investigate numerically the oxidation process in tube bundles, two additional phenomena
should be taken into account in the mathematical model.
The first of these is associated with a distribution of oxidant in the intertubular space.
The second phenomenon describes the oxidation of tubes, namely, the formation and growth of an oxide film on tubes 
in the cross--flow. As a rule, oxidant concentrations are very small and do not affect fluid dynamics. 
Under these conditions, the oxidant transport can be described using the conventional convection--diffusion model 
\citep{LandauLifshic1986}. Numerical simulation of flows around tube bundles in the presence of mass transfer
is performed \citep{li2005numerical} similarly to the calculation of the temperature.
Various oxidation models were developed \citep{young2016high} to describe mass transfer
in these flows.

In the present work, for predicting the oxidation process in a cross--flow around tube bundles,
a new model is developed to describe the growth of oxide films. 
It includes both the linear kinetics and classical parabolic kinetics of oxidation.
It is based (see, e.g., \cite{deal1965general,xu2011generalized}) on
the quasi--stationary approximation for a thin oxide film.
Using this model, a boundary value problem of mass transfer is formulated. 
The main feature of this problem is a nonlinear boundary condition of Robin--type for the oxidant concentration 
on tube surfaces. The local thickness of the film is considered as the desired value,
which, in particular, is explicitly included in the boundary condition for the oxidant concentration.

The paper is organized as follows. 
A new mathematical model of oxidation is developed in Section 2.
The fluid flow in the intertubular space is governed by the stationary incompressible Navier--Stokes equations. 
The flow around circular tubes arranged in in--line or staggered bundles is considered in the 2D formulation.
In Section 3, the computational algorithm for predicting hydrodynamics is given.
The numerical procedure employed in the work is based on using triangular grids and 
finite--element dscretization in space.
The numerical results of oxidation in the cross--flow around tube bundles are presented in Section 4. 
The oxidant transport is described by the unsteady convection--diffusion equation. 
To model oxidation processes, the combined model of linear and parabolic kinetics
is applied taking into account the growth of the oxide film.
The results of the work are summarized in Section 5.

\section{Mathematical model} 

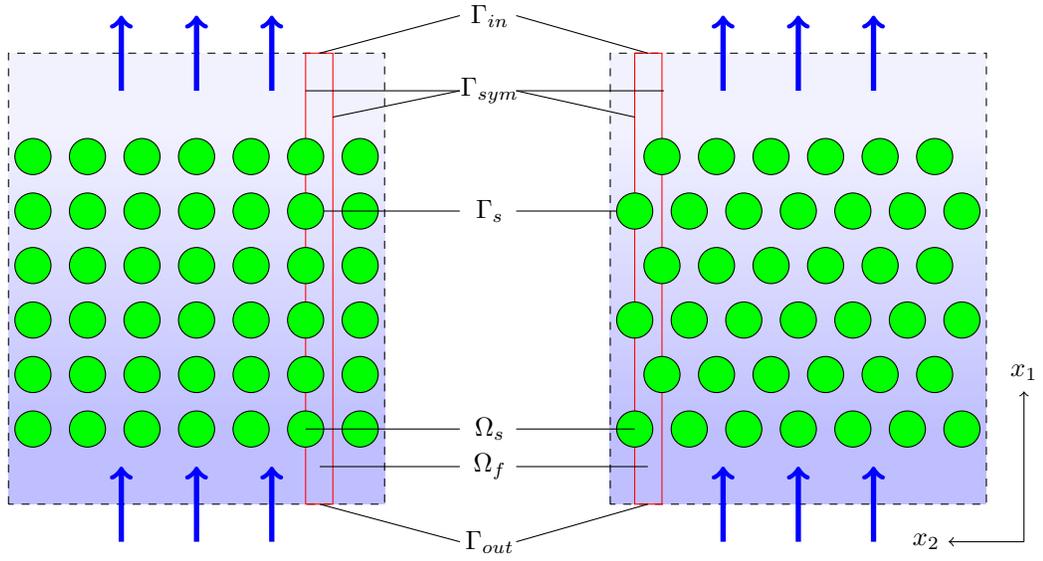
\begin{figure}[ht] 
  \begin{center}
    \begin{tikzpicture}
       \shade[top color=blue!5, bottom color=blue!5] (0,7) rectangle +(5,1);
       \shade[top color=blue!5, bottom color=blue!25] (0,3) rectangle +(5,4);
       \shade[top color=blue!25, bottom color=blue!25] (0,2) rectangle +(5,1);
       \draw [dashed] (0, 2) rectangle +(5,6);
       \draw [color=red] (4.3125-0.3625,2) rectangle +(0.3625,6);
       \foreach \y in {0,...,5} 
       \foreach \x in {0,...,6} {
  	\filldraw [fill=green,draw=black] (0.325+\x*0.725,3+\y*0.725) circle (0.24);
       }
       \draw [<-, line width=2, color=blue] (1.5,2.5) -- (1.5,1.5);
       \draw [<-, line width=2, color=blue] (2.5,2.5) -- (2.5,1.5);
       \draw [<-, line width=2, color=blue] (3.5,2.5) -- (3.5,1.5);
       \draw [<-, line width=2, color=blue] (1.5,8.5) -- (1.5,7.5);
       \draw [<-, line width=2, color=blue] (2.5,8.5) -- (2.5,7.5);
       \draw [<-, line width=2, color=blue] (3.5,8.5) -- (3.5,7.5);
       \shade[top color=blue!5, bottom color=blue!5] (8,7) rectangle +(5,1);
       \shade[top color=blue!5, bottom color=blue!25] (8,3) rectangle +(5,4);
       \shade[top color=blue!25, bottom color=blue!25] (8,2) rectangle +(5,1);
       \draw [dashed] (8, 2) rectangle +(5,6);
       \draw [color=red] (8.325,2) rectangle +(0.3625,6);
       \foreach \y in {0,...,2} 
	\foreach \x in {0,...,6} {
  	\filldraw [fill=green,draw=black] (8.325+\x*0.725,3+2*\y*0.725) circle (0.24);
       }
       \foreach \y in {0,...,2} 
	\foreach \x in {0,...,5} {
  	\filldraw [fill=green,draw=black] (8.325+0.3625+\x*0.725,3+0.725+2*\y*0.725) circle (0.24);
       }
       \draw [<-, line width=2, color=blue] (9.5,2.5) -- (9.5,1.5);
       \draw [<-, line width=2, color=blue] (10.5,2.5) -- (10.5,1.5);
       \draw [<-, line width=2, color=blue] (11.5,2.5) -- (11.5,1.5);
       \draw [<-, line width=2, color=blue] (9.5,8.5) -- (9.5,7.5);
       \draw [<-, line width=2, color=blue] (10.5,8.5) -- (10.5,7.5);
       \draw [<-, line width=2, color=blue] (11.5,8.5) -- (11.5,7.5);
       \draw [-] (4.5-0.3625,2.5) -- (6,2.5);
       \draw [-] (6.75,2.5) -- (8.5,2.5);
       \draw  (6.4,2.5) node {$\Omega_f$};
       \draw [-] (4.675-2*0.3625,3) -- (6,3);
       \draw [-] (6.75,3) -- (8.325,3);
       \draw  (6.4,3) node {$\Omega_s$};  
       \draw [-] (4.915-2*0.3625,5.9) -- (6,5.9);
       \draw [-] (6.75,5.9) -- (8.085,5.9);
       \draw  (6.4,5.9) node {$\Gamma_s$}; 
       \draw [-] (4.5-0.3625,2) -- (6,1.5);
       \draw [-] (6.75,1.5) -- (8.5,2);
       \draw  (6.4,1.5) node {$\Gamma_{out}$};
       \draw [-] (4.3125-0.3625,7.5) -- (6,7.5);
       \draw [-] (4.675-0.3625,7.15) -- (6,7.5);
       \draw [-] (8.325,7.15) -- (6.75,7.5);
       \draw [-] (8.7125,7.5) -- (6.75,7.5);
       \draw  (6.4,7.5) node {$\Gamma_{sym}$};
       \draw [-] (4.5-0.3625,8) -- (6,8.5);
       \draw [-] (6.75,8.5) -- (8.5,8.0);
       \draw  (6.4,8.5) node {$\Gamma_{in}$};  
       \draw [->] (13.5,1.5) -- (13.5,3.5);
       \draw  (13.5,3.75) node {$x_1$};  
       \draw [->] (13.5,1.5) -- (12.5,1.5);
       \draw  (12.2,1.5) node {$x_2$};  
    \end{tikzpicture}
    \caption{Cross--flow around tube bundles: in--line (left) and staggered (right) arrangement of tubes} 
   \label{f-1}
  \end{center}
\end{figure}

We consider a cross--flow around a bundle of circular tubes in the 2D formulation.
The sketch of the problem is given in Fig.~\ref{f-1}, where $x_1$ is the vertical (longitudinal) coordinate
and $x_2$ is the horizontal (transverse) coordinate. 
The tube rows of a bundle can be either in--line or staggered in the direction of the fluid velocity. 
They form a periodic structure of cylinders of the same circular cross-section.
The fluid flow contains oxygen, which oxidizes metal surfaces of the tubes.
Taking into account the symmetry of the flow, it is possible to work only with rectangular subdomains
shown in Fig.~\ref{f-1} by solid lines for both configurations.
The part $\Omega_f$ of these rectangles is occupied by the fluid, whereas the part $\Omega_s$ corresponds to the tubes.
The solid tube boundaries are denoted by $\Gamma_s$ and $\Gamma_{sym}$ stands for the symmetry boundaries.
The fluid with oxygen enters into the subdomains through the boundary $\Gamma_{in}$ and 
outflows on the boundary $\Gamma_{out}$. Mass transfer processes are considered in the computational domain $\Omega_f$.

\subsection{Hydrodynamic processes} 

To describe a cross-flow in a tube bundle, the incompressible Navier--Stokes equations are used:
\begin{equation}\label{1}
 \varrho \left ( \frac{\partial \bm{u} }{\partial t} + \bm{u} \cdot \nabla \bm{u} \right) 
 + \nabla p - \mu \nabla^2 \bm{u} = 0,
\end{equation} 
\begin{equation}\label{2}
 \nabla \cdot \bm{u} = 0, 
 \quad  \bm{x} \in \Omega_f, 
 \quad  t > 0,
\end{equation} 
where $\bm{u}(\bm{x},t)$ and $p(\bm{x},t)$ are the velocity and pressure of the fluid, respectively,
while $\mu > 0$ and $\rho > 0$ are, respectively, the viscosity and density assumed to be constant. 

Suitable boundary conditions are imposed on $\partial \Omega_f$.
These are the normal and tangential velocity components and/or components of the force applied
to the boundary, which are determined by the expression  $\bm \sigma \bm n$.
Here
\[
 \bm \sigma = \mu (\nabla \bm{u} + (\nabla \bm{u})^T )
\] 
is the viscous stress tensor and $\bm n$ stands for the outer unit normal to the boundary.
The uniform velocity profile with the value $\bar{u}$ is specified at the inlet:
\begin{equation}\label{3}
 \bm{u} \cdot \bm n = - \bar{u} ,
 \quad \bm{u} \times  \bm n = 0,
 \quad \bm x \in \Gamma_{in} .
\end{equation} 
The pressure $\bar{p}$ and condition for the absence of tangential forces are prescribed at the outlet:
\begin{equation}\label{4}
 p - \bm \sigma \bm n \cdot \bm n = \bar{p},
 \quad \bm \sigma \bm n \times \bm n = 0, 
 \quad \bm x \in \Gamma_{out} . 
\end{equation}
The no--slip and no--permeability conditions are imposed on the rigid walls of tubes:
\begin{equation}\label{5}
 \bm{u} \cdot \bm n = 0,
 \quad \bm{u} \times  \bm n = 0,
 \quad \bm x \in \Gamma_{s} .
\end{equation}
On the symmetry boundaries, we put the slip conditions:
\begin{equation}\label{6}
 \bm{u} \cdot \bm n = 0,
 \quad \bm \sigma \bm n \times \bm n = 0, 
 \quad \bm x \in \Gamma_{sym} .  
\end{equation} 

\subsection{Oxygen transport} 

We denote the oxygen concentration in the fluid by $c(\bm x, t)$, which is measured in particle number per unit volume. 
Assuming no fluid--phase reactions, the spatio--temporal evolution of the oxygen concentration is given by
\begin{equation}\label{7}
 \frac{\partial c }{\partial t} + \nabla (\bm{u} c) 
 - D \nabla^2 c = 0 ,
 \quad  \bm{x} \in \Omega_f, 
 \quad  t > 0,
\end{equation} 
where $D > 0$ is the solute diffusion coefficient, which is assumed to be scalar and constant. 

A known concentration of oxygen is specified at the inlet:
\begin{equation}\label{8}
 c = \bar{c},
 \quad \bm x \in \Gamma_{in} ,
\end{equation} 
where $\bar{c} > 0$ is constant.
We prescribe zero flux of the solute at the outlet and symmetry boundaries as follows:
\begin{equation}\label{9}
 D  \nabla c \cdot \bm{n} = 0,
 \quad \bm x \in \Gamma_{sym} \cup  \Gamma_{out} .
\end{equation} 

The key point in the description of the oxygen transport is related to modeling oxidation of metal surfaces.
In our case of small concentrations, it is natural to consider a thin layer of oxide.
To describe the growth of the oxide layer, various assumptions are employed \citep{young2016high}.

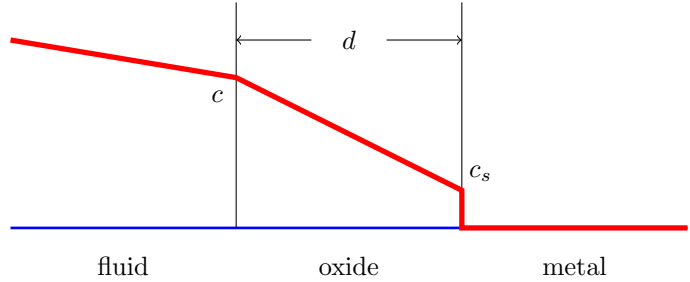
\begin{figure}[ht] 
  \begin{center}
    \begin{tikzpicture}
      \draw [-, line width=1, color=blue] (0,0) -- (9,0);
      \draw  (1.5,-0.5) node {fluid};
      \draw  (4.5,-0.5) node {oxide};
      \draw  (7.5,-0.5) node {metal};
      \draw [-] (3,0) -- (3,3);
      \draw [-] (6,0) -- (6,3);
      \draw [-, line width=2, color=red] (0,2.5) -- (3,2) -- (6,0.5) -- (6,0) -- (9,0);
      \draw [<-] (3,2.5) -- (4,2.5);
      \draw [->] (5,2.5) -- (6,2.5);
      \draw  (4.5,2.5) node {$d$};
      \draw  (2.75,1.75) node {$c$};
      \draw  (6.25,0.725) node {$c_s$};
    \end{tikzpicture}
    \caption{Sketch of the oxide film growth} 
   \label{f-2}
  \end{center}
\end{figure}

Following \cite{deal1965general,xu2011generalized}, we distinguish three stages of the oxygen transport:
\begin{enumerate}
 \item oxygen transport to the external surface of the oxide film;
 \item oxygen diffusion through the oxide film towards the oxide--metal interface;
 \item chemical reaction of oxidant with the reactive element at the oxide--metal interface.
\end{enumerate}
The second stage of the oxygen transport is associated with parabolic kinetics (the parabolic dependence
of the oxide film thickness on time), whereas the third one is linear kinetics (the linear time--evolution).

Let us denote the oxygen concentration at the fluid--oxide boundary as $c$,
and at the oxide--metal interface as $c_s$ (see Fig.~\ref{f-2}).
Here $d$ is the oxide film thickness.

Taking into account the fact that the thickness of the oxide film is small, we can treat
the film growth as a quasi--stationary process. In this case, the oxidant flux through the film
is continuous. In addition, this flux is continuous both at the fluid--oxide boundary and
at the oxide--metal interface due to the mass conservation law.

The mass flux at the fluid--oxide boundary from the liquid side is
\begin{equation}\label{10}
 q_f = - D  \nabla c \cdot \bm{n}, 
\end{equation} 
where $\bm{n}$ is the outer unit normal for the fluid domain.

For the mass flux in the oxide film, we have
\begin{equation}\label{11}
 q_o = D_0 \frac{c - c_s}{d} ,
\end{equation} 
where $D_0$ is the diffusion coefficient of oxygen in the oxide film.

Oxygen arrived at the oxide--metal interface participates in the chemical reaction of
oxidation. The mass of oxygen involved in oxidation is evaluated as
\begin{equation}\label{12}
 q_m = k c_s ,
\end{equation} 
where $k$ is the oxidation rate (the first--order chemical reaction).

For the quasi--stationary regime, we have
\begin{equation}\label{13}
 q_f = q_o ,
 \quad  q_o = q_m .
\end{equation}
From the second equality in (\ref{13}), in view of (\ref{11}), (\ref{12}), we can eliminate $c_s$.
Then, the first equality in (\ref{13}), in view of (\ref{10}), (\ref{11}), makes possible to
formulate the boundary condition for the concentration of oxygen at the fluid boundary:
\begin{equation}\label{14}
 D  \nabla c \cdot \bm{n} + \frac{k D_0}{D_0 + k d} c = 0, 
 \quad \bm x \in \Gamma_{s}, 
\end{equation}
which depends on the thickness of the oxide film.

The time--variation of the thickness of the oxide film at the boundary point is evaluated via the reacted oxidant:
\begin{equation}\label{15}
 \frac{\partial  d}{\partial t} \varrho_0 = \frac{k D_0}{D_0 + k d} c ,
 \quad \bm x \in \Gamma_{s} ,
\end{equation} 
where  $\varrho_0$ is the density of the oxide film.
In the dependence (\ref{15}) for the local thickness of the oxide film, two limiting cases can be distinguished.
For a small thickness of the oxide film ($d \ll D_0 / k$), linear kinetics is realized:
\[
 \frac{\partial  d}{\partial t} \varrho_0 \approx k c .
\]  
For large values of the thickness ($d \gg D_0 / k$), we have parabolic kinetics:
\[
 \frac{\partial  d}{\partial t} \varrho_0 \approx \frac{D_0}{d}  c .
\]  

To consider the dynamic processes, the system of equations (\ref{1}), (\ref{2}), (\ref{7}),
in addition to the boundary conditions (\ref{3})--(\ref{6}), (\ref{8}), (\ref{9}), (\ref{14}), 
is supplemented with equation (\ref{15}) describing the time--evolution of the thickness of the oxide film. 
Also, we must specify the initial conditions for the velocity
\[
 \bm{u}(\bm x, 0) = \bm{u}_0(\bm x), 
 \quad \bm x \in \Omega_f , 
\]
as well as for the oxygen concentration and the thickness of the oxide film, respectively:
\begin{equation}\label{16}
 c(\bm x, 0) = c_0(\bm x), 
 \quad \bm x \in \Omega_f . 
\end{equation} 
\begin{equation}\label{17}
 d(\bm x, 0) = d_0(\bm x), 
 \quad \bm x \in \Gamma_s. 
\end{equation} 

It seems natural to simplify the hydrodynamic problem and assume that the fluid flow is steady--state.
In view of this, instead of (\ref{1}), we solve the equation
\begin{equation}\label{18}
 \varrho \bm{u} \cdot \nabla \bm{u}  
 + \nabla p - \mu \nabla^2 \bm{u} = 0 .
\end{equation}
Thus, modeling of the oxidation process is conducted via solving the initial--boundary value problem
(\ref{2})--(\ref{9}), (\ref{14})--(\ref{18}).

\subsection{Dimensionless problem} 

Let us formulate the above problem in the dimensionless form, using for the dimensionless velocity, pressure and concentration the same notation as for the dimensional ones.
As the reference values, we take the diameter of tubes $l$, the inlet velocity value $\bar{u}$
and the inlet concentration $\bar{c}$.
Then equation (\ref{18}) takes the form
\begin{equation}\label{19}
 \bm{u} \cdot \nabla \bm{u}  
 + \nabla p - \frac{1}{\mathrm{Re}} \nabla^2 \bm{u} = 0 ,
\end{equation}
where
\[
 \mathrm{Re} = \frac{\varrho l \bar{u} }{\mu}
\]
is the Reynolds number. The boundary condition  (\ref{3}) takes the form
\begin{equation}\label{20}
 \bm{u} \cdot \bm n = - 1,
 \quad \bm{u} \times  \bm n = 0,
 \quad \bm x \in \Gamma_{in} .
\end{equation}
Taking into account the fact that the pressure is determined to within a constant, rewrite (\ref{4}) in the form
\begin{equation}\label{21}
 p - \bm \sigma \bm n \cdot \bm n = 0,
 \quad \bm \sigma \bm n \times \bm n = 0, 
 \quad \bm x \in \Gamma_{out} . 
\end{equation} 
In the dimensionless variables, we get
\[
 \bm \sigma = \nabla \bm{u} + (\nabla \bm{u})^T .
\] 
The boundary conditions (\ref{5}), (\ref{6}) remain unchanged.

The equation (\ref{7}) in the dimensionless form is written as
\begin{equation}\label{22}
 \frac{\partial c }{\partial t} + \nabla (\bm{u} c) 
 - \frac{1}{\mathrm{Pe}}  \nabla^2 c = 0 ,
 \quad  \bm{x} \in \Omega_f, 
 \quad  t > 0,
\end{equation} 
where
\[
 \mathrm{Pe} =  \frac{l \bar{u} }{D} 
\] 
is the Peclet number.

From (\ref{8}), we have
\begin{equation}\label{23}
 c = 1,
 \quad \bm x \in \Gamma_{in} ,
\end{equation} 
and the boundary condition (\ref{9}) takes the form
\begin{equation}\label{24}
 \nabla c \cdot \bm{n} = 0,
 \quad \bm x \in \Gamma_{sym} \cup  \Gamma_{out} .
\end{equation} 

The boundary condition (\ref{14}) in the dimensionless form is written as following:
\begin{equation}\label{25}
 \nabla c \cdot \bm{n} + \left ( \frac{1}{\mathrm{Sh}_1} + \frac{1}{\mathrm{Sh}_2} d \right )^{-1}  c = 0,
 \quad \bm x \in \Gamma_{s}.
\end{equation}
Here $\mathrm{Sh}_1$ is the Sherwood number that corresponds to the oxidation process based on linear kinetics:
\[
 \mathrm{Sh}_1 = \frac{k l}{D} .
\] 
For the Sherwood number corresponding to the approximation of parabolic kinetics, we have
\[
 \mathrm{Sh}_2 = \frac{l}{D} \frac{D_0}{\bar{d}} ,
 \quad \bar{d} = \frac{D \bar{c}}{\varrho_0 \bar{u}} .
\]
Using the above reference value $\bar{d}$ for the thickness of the oxide film, 
from (\ref{15}), we arrive at the dimensionless equation
\begin{equation}\label{26}
 \frac{\partial  d}{\partial t}  = \left ( \frac{1}{\mathrm{Sh}_1} + \frac{1}{\mathrm{Sh}_2} d \right )^{-1}  c,
 \quad \bm x \in \Gamma_{s} .
\end{equation} 

Thus, the problem under consideration is characterized by two mass transfer parameters, namely,
$\mathrm{Re}$ and $\mathrm{Pe}$, as well as the oxidation process parameters $\mathrm{Sh}_1$ and $\mathrm{Sh}_2$.
It should be noted that the boundary conditions for the oxidant concentration are non--linear
(see (\ref{25}), (\ref{26})).

\section{Hydrodynamic processes} 

The numerical solution of the 2D hydrodynamic problem of a cross-flow around tube bundles is obtained 
on the basis of finite--element discretization in space.

\subsection{Computational domain and grids} 

The triangulation of the computational domain $\Omega_f$ is performed using the Gmsh grid generator 
(website gmsh.info, \cite{Gmsh}). Scripts for preparing geometries of both configurations are written in the Python programming language.

\begin{sidewaysfigure}
  \begin{center}
    \includegraphics[width=1\linewidth] {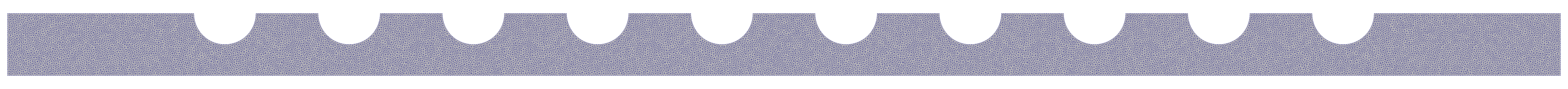}

	\vspace{5mm} a \vspace{5mm} 

    \includegraphics[width=1\linewidth] {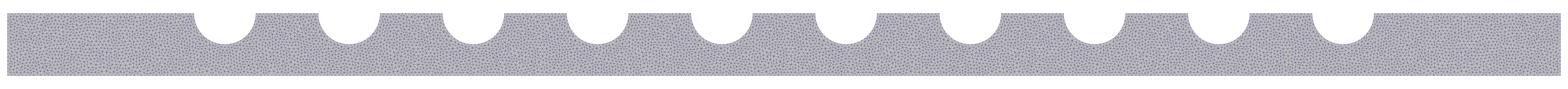}

	\vspace{5mm} b \vspace{5mm} 

    \includegraphics[width=1\linewidth] {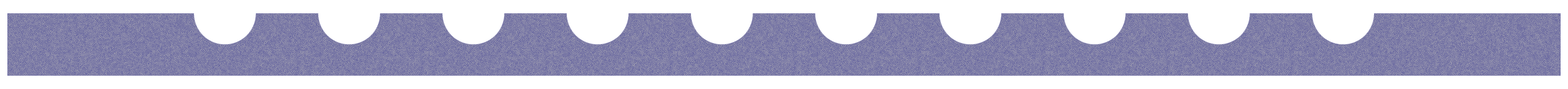}

	\vspace{5mm} c \vspace{5mm} 

	\caption{Computational grids for the in--line configuration: 
	         a --- basic (medium) grid (23,493 vertices and 45,328 cells),
                 b --- coarse grid (8,370 vertices and 15,764 cells), 
                 c --- fine grid (71,326 vertices and 139,760 cells)}
	\label{f-3}
  \end{center}
\end{sidewaysfigure}

\begin{sidewaysfigure}
  \begin{center}
    \includegraphics[width=1\linewidth] {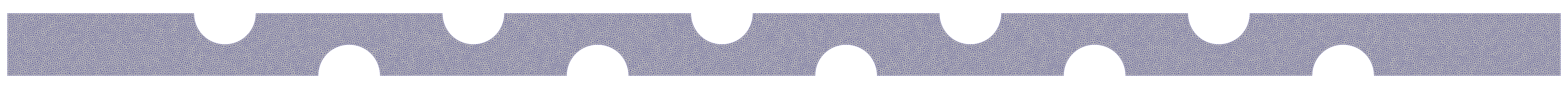}

	\vspace{5mm} a \vspace{5mm} 

    \includegraphics[width=1\linewidth] {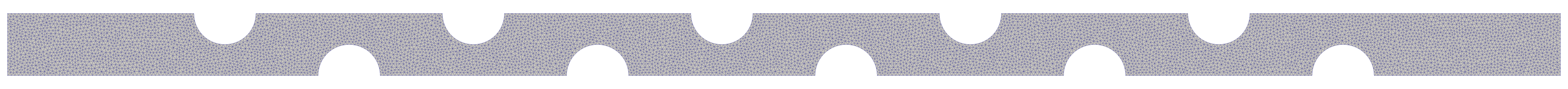}

	\vspace{5mm} b \vspace{5mm} 

    \includegraphics[width=1\linewidth] {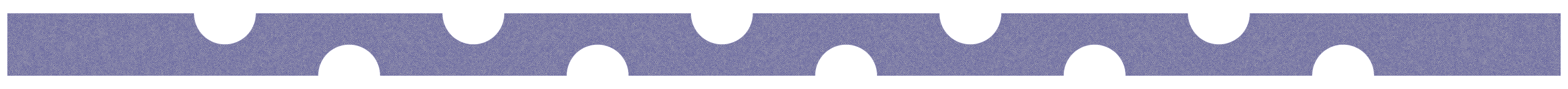}

	\vspace{5mm} c \vspace{5mm} 

	\caption{Computational grids for the staggered configuration: 
	         a --- basic (medium) grid (23,511 vertices and 45,366 cells),
                 b --- coarse grid (8,362 vertices and 15,750 cells), 
                 c --- fine grid (71,246 vertices and 139,600 cells)}
	\label{f-4}
  \end{center}
\end{sidewaysfigure}

In the in--line configuration, we consider 10 circular tubes with the dimensionless diameter equals 1 
arranged in the bundle with the longitudinal and transverse pitches (measured between tube centers)
equal 2. The grid independence study was conducted using the sequence of refined grids presented in Fig.~\ref{f-3}. 
The staggered configuration has the same pitches, the corresponding computational grids are shown in Fig.~\ref{f-4}.

\subsection{Computational algorithm} 

The hydrodynamic problem was solved separately from mass transfer. The finite--element discretization
\citep{gresho200incompressible} is based on the variational formulation
for the boundary value problem (\ref{19}), (\ref{2}), (\ref{5}), (\ref{6}), (\ref{20}), (\ref{21}). 
For the velocity $\bm u$, we define the function space $\bm V$ ($\bm u \in \bm V$):
\[
\begin{split}
 \bm V = \{ \bm u \in \bm H^1(\Omega_f) : \  & \bm{u} \cdot \bm n = - 1, \
 \bm{u} \times  \bm n = 0 \ \mathrm{on} \ \Gamma_{in} , 
 \\ & \bm{u} = 0 \ \mathrm{on} \ \Gamma_{s}, \ \bm{u} \cdot \bm n = 0 \
 \ \mathrm{on} \ \Gamma_{sym} \} .
\end{split} 
\] 
For test functions $\bm v \in \hat{\bm V}$, we have
\[
 \hat{\bm V} = \{ \bm v \in \bm H^1(\Omega_f) : \ \bm{v} = 0 \ \mathrm{on} \ \Gamma_{in} , \ 
 \bm{v} = 0 \ \mathrm{on} \ \Gamma_{s}, \ \bm{v} \cdot \bm n = 0 \
 \ \mathrm{on} \ \Gamma_{sym} \} .
\]
For the pressure $p$ and the corresponding test function $q$, we set $p, q \in Q$, where
\[
 Q = \{ q \in L_2(\Omega_f) : \ q = 0 \ \mathrm{on} \  \Gamma_{out} \} .
\]
Multiply equation (\ref{17}) by $\bm v$ and integrate it over the whole domain. 
The similar transformation of equation (\ref{2}) is performed using $q$. 
Taking into account the boundary conditions (\ref{6}), (\ref{20}), (\ref{21}), we obtain the system of equations:
\begin{equation}\label{27}
 a(\bm u, \bm v) - b(\bm v, p) = 0 \ \forall \bm v \in \hat{\bm V} ,
\end{equation} 
\begin{equation}\label{28}
 b(\bm u, q) = 0 \ \forall q \in Q
\end{equation} 
for the desired $\bm u \in \bm V$, $p \in Q$. 
Here
\[
 a(\bm u, \bm v) := \int_{\Omega_f} (\bm{u} \cdot \nabla \bm{u}) \cdot \bm v \, d \bm x +  
 \frac{1}{\mathrm{Re}}\int_{\Omega_f} \nabla \bm u \cdot \nabla \bm v \, d \bm x , 
\] 
\[
 b(\bm v, q) := \int_{\Omega_f}(\nabla \cdot \bm v) q \, d \bm x . 
\] 

To define the finite--element discretization, we choose finite--dimensional subspaces
$\bm V_h \subset \bm V$, $\hat{\bm V}_h \subset \hat{\bm V}$ and $Q_h \subset Q$ for the approximate solution and test functions. Here we use the Taylor--Hood $P_2-P_1$ finite element \citep{taylor1973numerical}.
It consists of a continuous $P_2$ Lagrange element for the velocity components 
and a continuous $P_1$ Lagrange element for the pressure field.

To solve the nonlinear variational problem, the iterative Newton method is applied. 
The computational implementation is based on the FEniCS platform for solving partial differential equations 
(website fenicsproject.org, \cite{LoggMardalEtAl2012a,AlnaesBlechta2015a}).
The convergence of the iterative Newton method for various values of the Reynolds number
is presented in Table~\ref{t-1}.

\begin{table}[htp]
\caption{Convergence of the iterative process}
\label{t-1}
  \begin{center}
  \begin{tabular}{|c|c|c|c|}
  \hline
   Re & Iteration & Absolute residual  & Relative residual \\
      &           & in--line/staggered & in--line/staggered \\
  \hline
     & 1 & 1.427e-01/1.874e-01  & 1.858e-02/2.439e-02 \\
  10 & 2 & 9.574e-03/3.373e-02  & 1.246e-03/4.392e-03 \\
     & 3 & 6.448e-06/4.395e-05  & 8.395e-07/5.722e-06 \\
     & 4 & 1.885e-11/2.581e-10  & 2.454e-12/3.360e-11 \\
  \hline
     & 1 & 1.427e-01/1.874e-01  & 1.858e-02/2.439e-02 \\
  50 & 2 & 3.531e-02/8.926e-02  & 4.597e-03/1.162e-02 \\
     & 3 & 7.390e-04/8.966e-03  & 9.621e-05/1.167e-03 \\
     & 4 & 4.896e-06/2.646e-04  & 6.374e-07/3.445e-05 \\
  \hline
     & 1 & 1.427e-01/1.874e-01  & 1.858e-02/2.439e-02 \\
 150 & 2 & 4.975e-02/9.916e-02  & 6.477e-03/1.291e-02 \\
     & 3 & 3.886e-03/9.707e-02  & 5.059e-04/1.264e-02 \\
     & 4 & 3.283e-04/4.778e-02  & 4.274e-05/6.220e-03 \\
  \hline
  \end{tabular}
  \end{center}
\end{table} 

\subsection{Numerical results for the stationary hydrodynamic problem} 

\begin{sidewaysfigure}
  \begin{center}
    \includegraphics[width=1\linewidth] {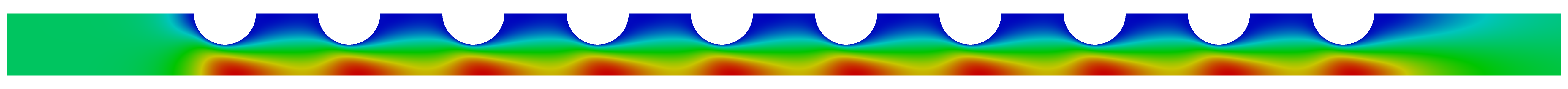}

	\vspace{5mm} a \vspace{5mm} 

    \includegraphics[width=1\linewidth] {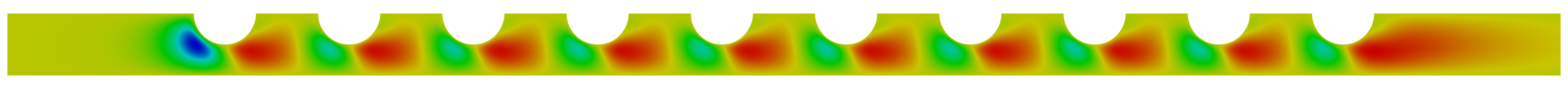}

	\vspace{5mm} b \vspace{5mm} 

    \includegraphics[width=1\linewidth] {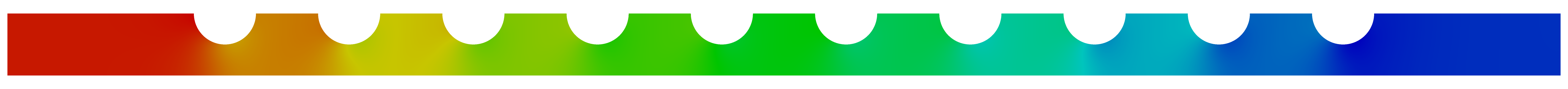}

	\vspace{5mm} c \vspace{5mm} 

	\caption{Velocity and pressure for the in--line configuration, $\mathrm{Re} =10$: 
	         a --- the longitudinal velocity component $u_1$,
                 b --- the transverse velocity component $u_2$, 
                 c --- the pressure}
	\label{f-5}
  \end{center}
\end{sidewaysfigure}

\begin{sidewaysfigure}
  \begin{center}
    \includegraphics[width=1\linewidth] {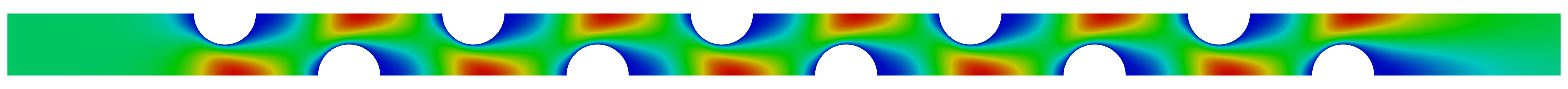}

	\vspace{5mm} a \vspace{5mm} 

    \includegraphics[width=1\linewidth] {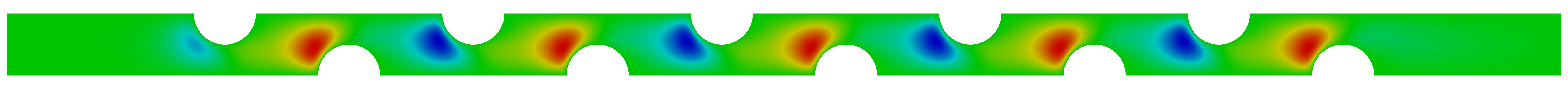}

	\vspace{5mm} b \vspace{5mm} 

    \includegraphics[width=1\linewidth] {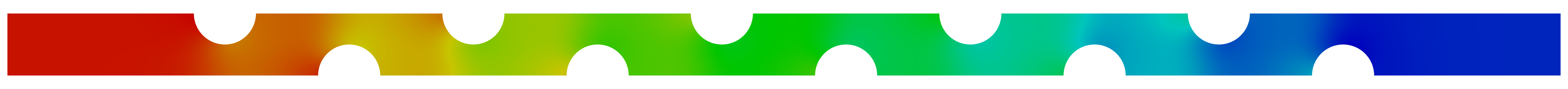}

	\vspace{5mm} c \vspace{5mm} 

	\caption{Velocity and pressure for the staggered configuration, $\mathrm{Re} =10$:
		 a --- the longitudinal velocity component $u_1$,
                 b --- the transverse velocity component $u_2$, 
                 c --- the pressure}
        \label{f-6}
  \end{center}
\end{sidewaysfigure}

We start our numerical study with $\mathrm{Re} =10$. The velocity components and pressure calculated 
on the basic grid for the in--line configuration are shown in Fig.~\ref{f-5}.
Similar data for the staggered configuration are given in Fig.~\ref{f-6}.

\begin{sidewaysfigure}   
  \begin{center}

    \includegraphics[width=1.1\linewidth] {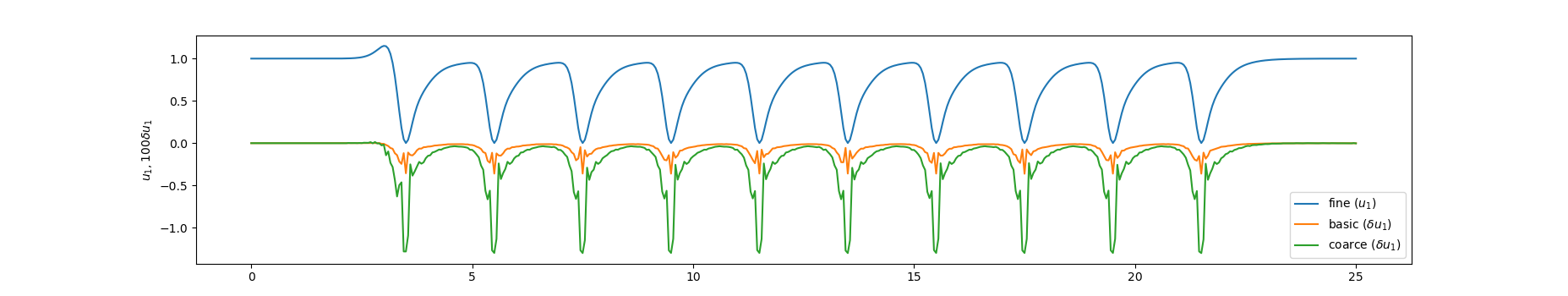}

	\vspace{5mm} a \vspace{5mm} 

    \includegraphics[width=1.1\linewidth] {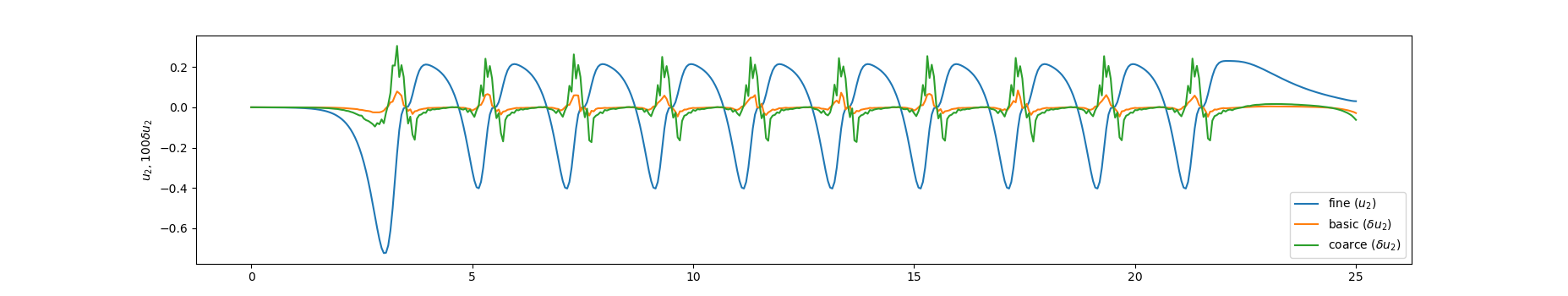}

	\vspace{5mm} b \vspace{5mm} 

    \includegraphics[width=1.1\linewidth] {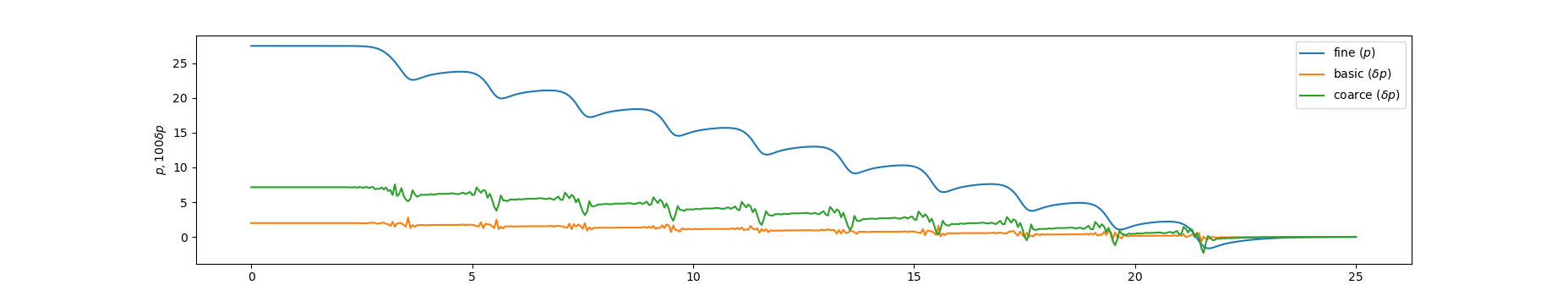}

	\vspace{5mm} c \vspace{5mm} 

	\caption{Velocity components and the pressure along the midline of the computational domain
	         for the in--line configuration presented as the solution on the fine grid and 
	         the normalized deviations from it for the basic and coarse grids: 
		 a --- the longitudinal velocity component $u_1$,
                 b --- the transverse velocity component $u_2$, 
                 c --- the pressure}
	\label{f-7}
  \end{center}
\end{sidewaysfigure}

\begin{sidewaysfigure}   
  \begin{center}

    \includegraphics[width=1.1\linewidth] {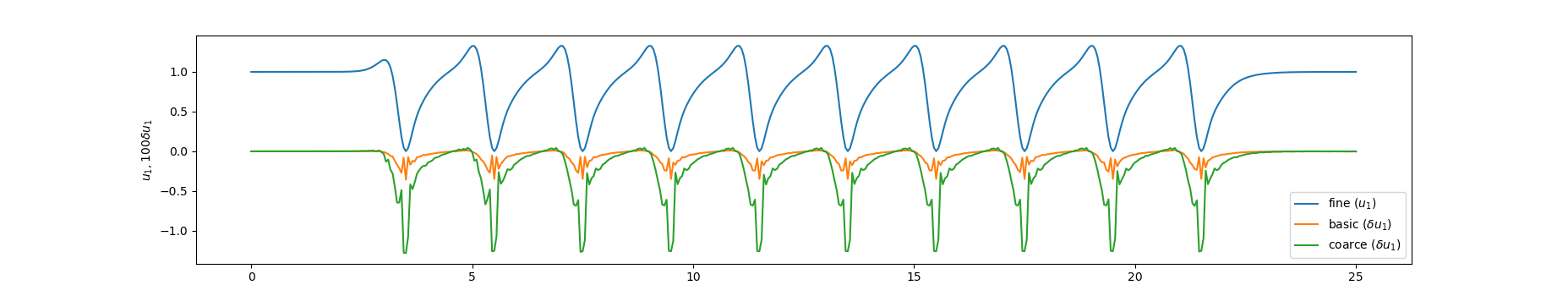}

	\vspace{5mm} a \vspace{5mm} 

    \includegraphics[width=1.1\linewidth] {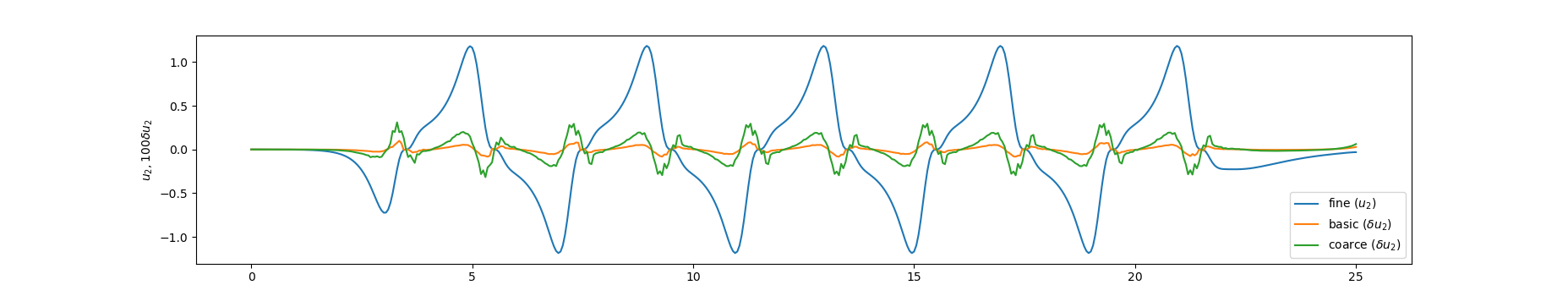}

	\vspace{5mm} b \vspace{5mm} 

    \includegraphics[width=1.1\linewidth] {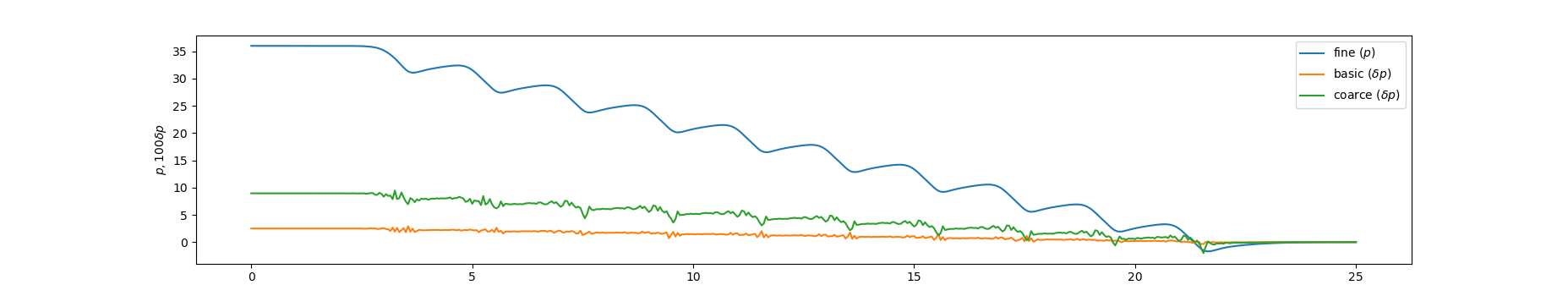}

	\vspace{5mm} c \vspace{5mm} 

	\caption{Velocity components and the pressure along the midline of the computational domain
	         for the staggered configuration presented as the solution on the fine grid and
		      the normalized deviations from it for the basic and coarse grids: 
		 a --- the longitudinal velocity component $u_1$,
                 b --- the transverse velocity component $u_2$, 
                 c --- the pressure}
	\label{f-8}
  \end{center}
\end{sidewaysfigure}

The grid convergence of the numerical solution demonstrates Fig.~\ref{f-7} for the in-line configuration 
and Fig.~\ref{f-8} for the staggered configuration, respectively. In these figures,
the velocity components and the pressure ($u_1, u_2, p$) are given versus $x_1$ 
along the midline of the computational domain. These main plots are obtained on the fine grid. 
In addition, there is depicted the deviation ($\delta u_1, \delta u_2, \delta p$) from this finest solution.
These two deviations corresponding to the coarse and basic (medium) grids are normalized (multiplied by 100)
in order to make this visualization more evident. These figures indicate a good accuracy of the numerical results
obtained on the fine grid.

\begin{sidewaysfigure}   
  \begin{center}

    \includegraphics[width=1.1\linewidth] {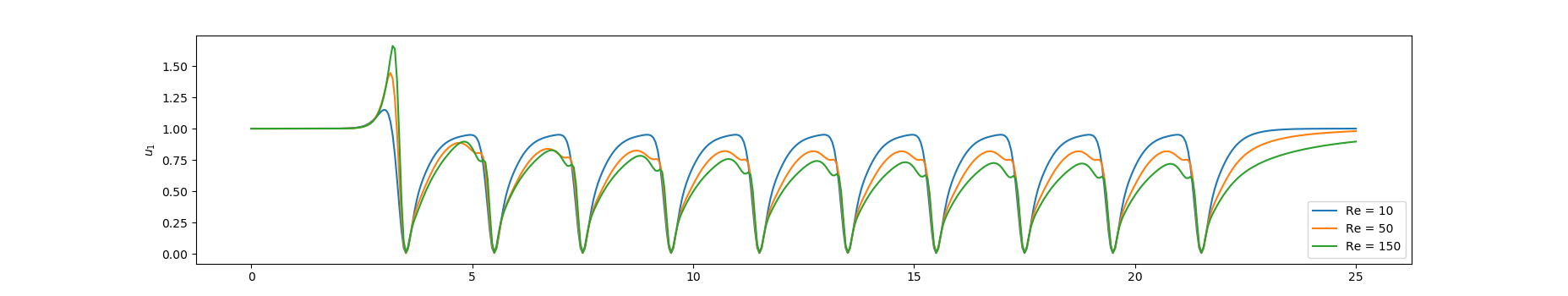}

	\vspace{5mm} a \vspace{5mm} 

    \includegraphics[width=1.1\linewidth] {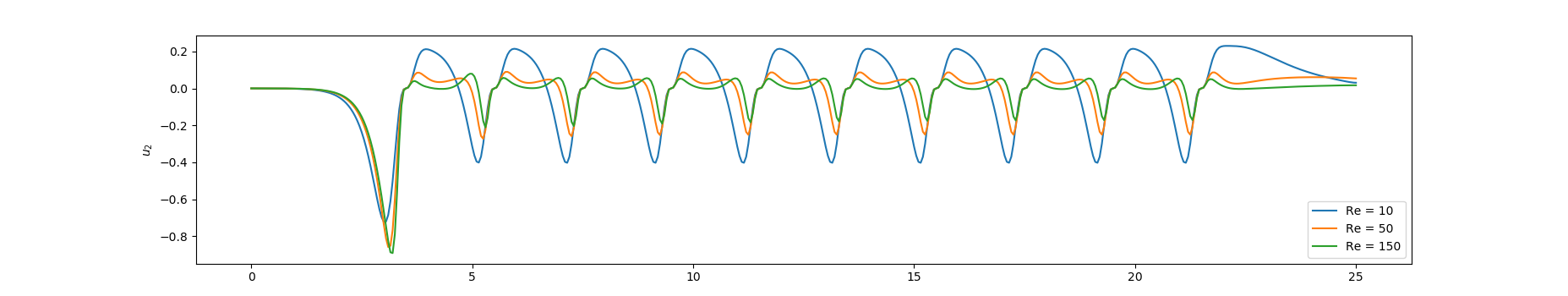}

	\vspace{5mm} b \vspace{5mm} 

    \includegraphics[width=1.1\linewidth] {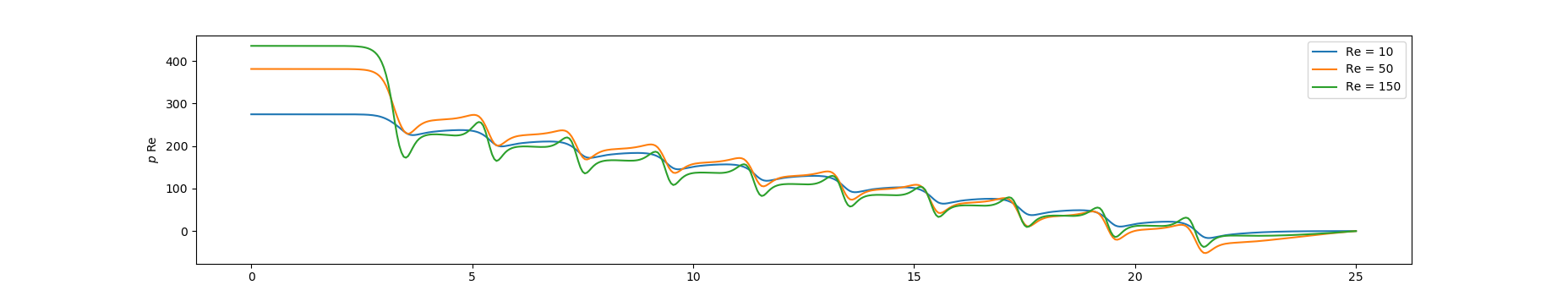}

	\vspace{5mm} c \vspace{5mm} 

	\caption{Velocity components and the pressure along the midline of the computational domain
	         for the in--line configuration at $\mathrm{Re} = 10, 50, 150$:
	         a --- the longitudinal velocity component $u_1$,
                 b --- the transverse velocity component  $u_2$, 
                 c --- the pressure}
	\label{f-9}
  \end{center}
\end{sidewaysfigure}

\begin{sidewaysfigure}   
  \begin{center}

    \includegraphics[width=1.1\linewidth] {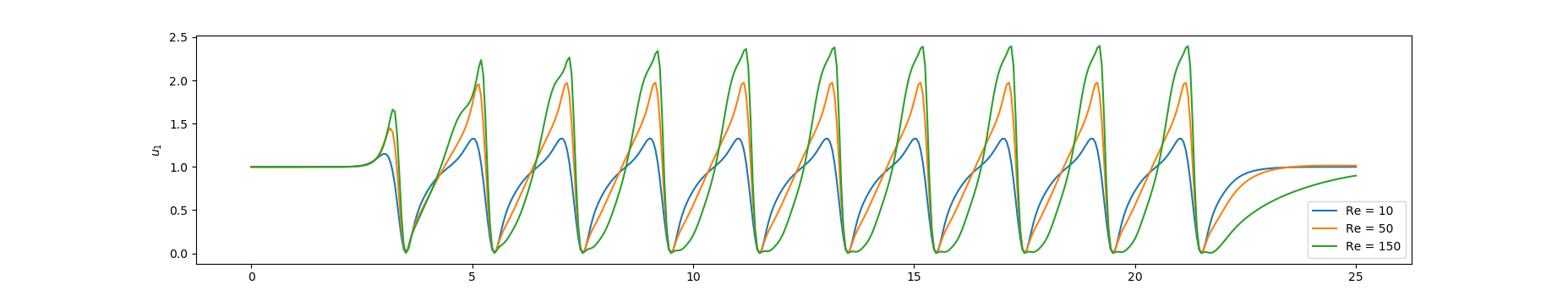}

	\vspace{5mm} a \vspace{5mm} 

    \includegraphics[width=1.1\linewidth] {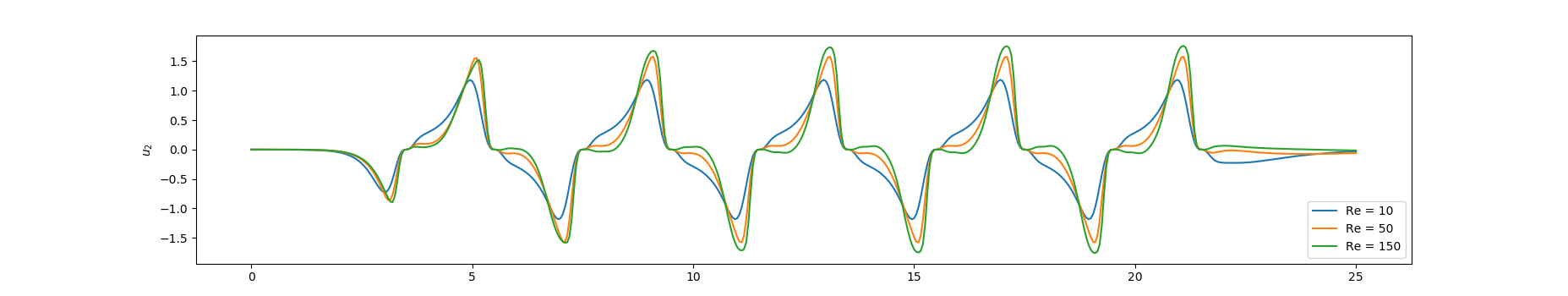}

	\vspace{5mm} b \vspace{5mm} 

    \includegraphics[width=1.1\linewidth] {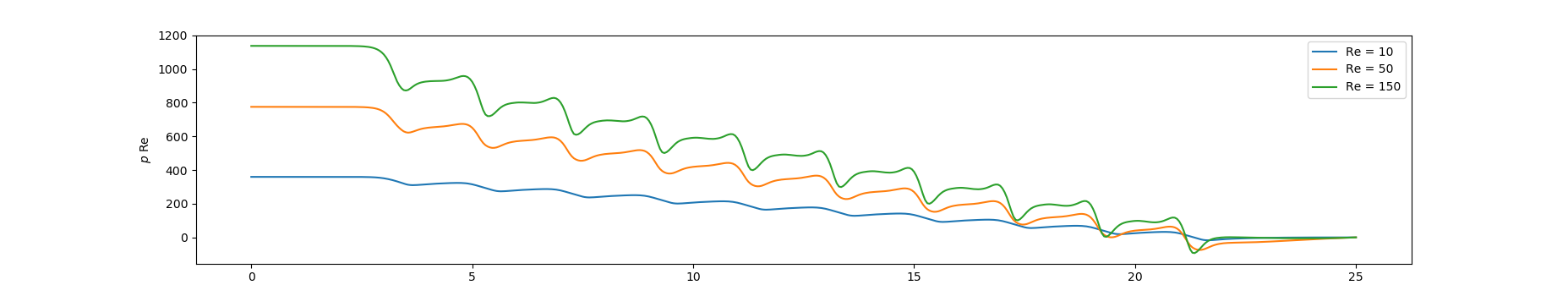}

	\vspace{5mm} c \vspace{5mm} 

    \caption{Velocity components and the pressure along the midline of the computational domain
	     for the staggered configuration at $\mathrm{Re} = 10, 50, 150$:
	     a --- the longitudinal velocity component $u_1$,
             b --- the transverse velocity component  $u_2$, 
             c --- the pressure}
	\label{f-10}
  \end{center}
\end{sidewaysfigure}

\begin{sidewaysfigure}   
  \begin{center}

    \includegraphics[width=1.1\linewidth] {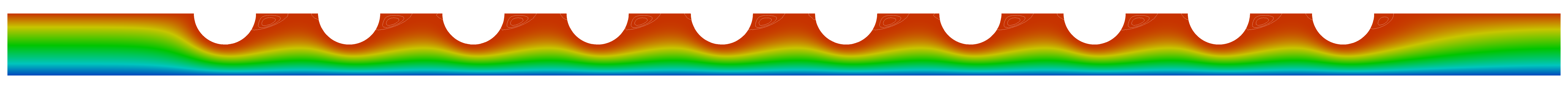}

	\vspace{5mm} a \vspace{5mm} 

    \includegraphics[width=1.1\linewidth] {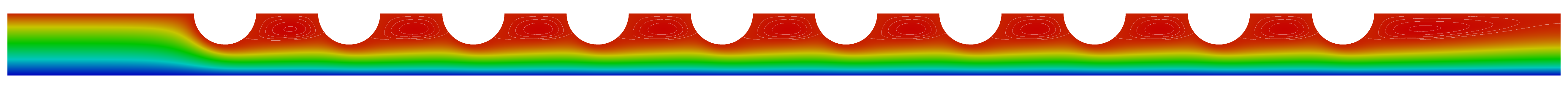}

	\vspace{5mm} b \vspace{5mm} 

    \includegraphics[width=1.1\linewidth] {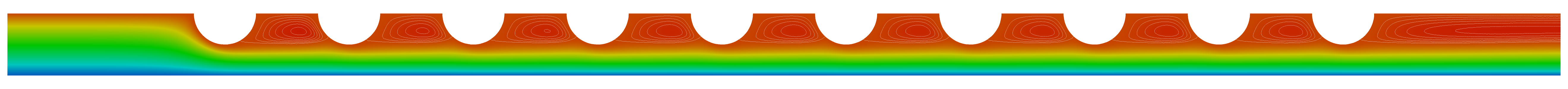}

	\vspace{5mm} c \vspace{5mm} 

	\caption{Streamlines for the in--line configuration at various Reynolds numbers: 
	         a --- $\mathrm{Re} = 10$,
                 b --- $\mathrm{Re} = 50$, 
                 c --- $\mathrm{Re}= 150$}
	\label{f-11}
  \end{center}
\end{sidewaysfigure}

\begin{sidewaysfigure}   
  \begin{center}

    \includegraphics[width=1.1\linewidth] {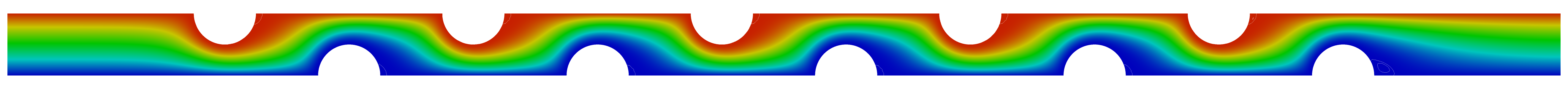}

	\vspace{5mm} a \vspace{5mm} 

    \includegraphics[width=1.1\linewidth] {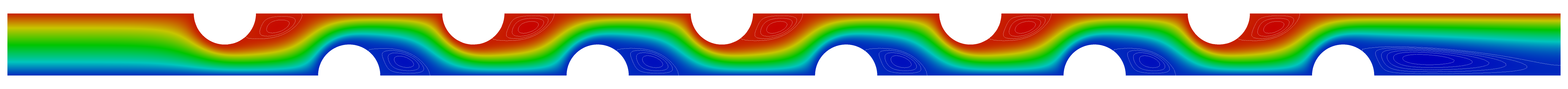}

	\vspace{5mm} b \vspace{5mm} 

    \includegraphics[width=1.1\linewidth] {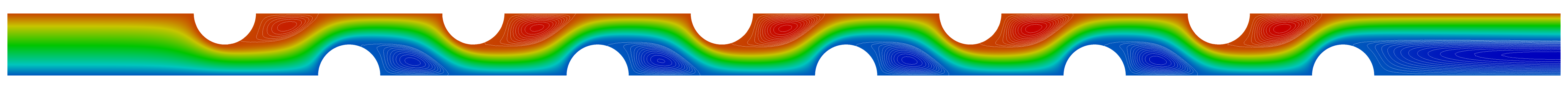}

	\vspace{5mm} c \vspace{5mm}
	
    \caption{Streamlines for the staggered configuration at various Reynolds numbers: 
	     a --- $\mathrm{Re} = 10$,
             b --- $\mathrm{Re} = 50$, 
             c --- $\mathrm{Re}= 150$}
	\label{f-12}
  \end{center}
\end{sidewaysfigure}

The effect of the Reynolds number on the flow is shown in Fig.~\ref{f-9} for the in--line configuration
and in Fig.~\ref{f-10} for the staggered configuration, respectively. For $\mathrm{Re} < 1$, in fact, 
we can restrict ourselves to the Stokes approximation (do not take into account the convective terms 
in equation (\ref{19})). To show flow patterns for these 2D stationary flows, we employ the streamfunction
$\psi$ defined by the relation
\[
 \bm u = \left (\frac{\partial \psi }{\partial x_2}, - \frac{\partial \psi }{\partial x_1} \right ) .
\] 
It is evaluated from the known velocity using the equation
\[
 - \nabla^2 \psi = \omega , 
 \quad \bm x \in \Omega_f, 
\] 
where
\[
 \omega = \frac{\partial u_2 }{\partial x_1} - \frac{\partial u_1 }{\partial x_2} 
\]
and the corresponding Dirichlet boundary conditions are specified on $\partial \Omega_f$. 
Figures~\ref{f-11} and \ref{f-12} present streamlines at different Reynolds numbers for the in--line and
staggered configurations, respectively.

\section{Oxidation process} 

The oxidation process is modeled via solving the unsteady convection--diffusion equation
with the corresponding conditions on the tube surfaces.

\subsection{Solution of the mass transfer problem} 

The unsteady problem of oxygen transport (\ref{20})--(\ref{26}) with the initial conditions (\ref{16}), (\ref{17})
is solved numerically using the Lagrangian finite elements $P_1$. Define
\[
 S = \{ s \in H^1(\Omega_f) : \ s = 1 \ \mathrm{on} \  \Gamma_{in} \} ,
\]  
\[
 \hat{S} = \{ s \in H^1(\Omega_f) : \ s = 0 \ \mathrm{on} \  \Gamma_{in} \} .
\]
The concentration $c \in S$ is obtained from the equation
\begin{equation}\label{29}
\left (\frac{\partial c}{\partial t}, s \right) + e(c,s) =  
 \left (\frac{\partial d}{\partial t} ,s \right )_s
 \quad \forall s \in \hat{S} ,   
\end{equation} 
where
\[
  e(c,s) := - \int_{\Omega_f} c \bm u \cdot \nabla  s \, d \bm x 
  + \frac{1}{\mathrm{Pe}} \int_{\Omega_f} \nabla c \cdot \nabla  s \, d \bm x 
  + \int_{\Gamma_{out}} (\bm u \cdot \bm n) c s \, d \bm x ,
\] 
\[
  (\varphi,s)_s := - \int_{\Gamma_{s}} \varphi   s \, d \bm x .
\] 
For $d \in G = L_2(\Gamma_{s})$, from (\ref{26}), we have
\begin{equation}\label{30}
 \left (\frac{\partial d}{\partial t} , g \right )_s -
 \left (\left ( \frac{1}{\mathrm{Sh}_1} + \frac{1}{\mathrm{Sh}_2} d \right )^{-1}  c, g \right )_s = 0,
 \quad g \in G .  
\end{equation} 

For time--stepping, we employ the Crank--Nicolson scheme of second order \cite{Samarskii,Ascher2008}. 
Let $\tau$ be a step--size of a uniform grid in time such that 
$c^n = c(t^n), t^n = n\tau, \ n = 0, 1, ...$. 
For equation (\ref{29}),  we apply the following two--level scheme:
\begin{equation}\label{31}
 \left (\frac{c^{n+1} - c^n}{\tau }, s \right) + e \left (\frac{c^{n+1} - c^n}{2} ,s\right ) =  
 \left (\frac{d^{n+1} - d^n}{\tau } ,s \right )_s 
 \quad n = 0, 1, ... . 
\end{equation} 
Similarly, for (\ref{30}), we have
\begin{equation}\label{32}
 \left (\frac{d^{n+1} - d^n}{\tau }  , g \right )_s -
 \left (\left ( \frac{1}{\mathrm{Sh}_1} + \frac{1}{\mathrm{Sh}_2} \frac{d^{n+1} + d^n}{2} \right )^{-1}  \frac{c^{n+1} + c^n}{2}, g \right )_s = 0, 
\end{equation}
for the given initial conditions (see (\ref{16}), (\ref{17})).
We confine ourselves to the case of the homogeneous initial conditions:
\[
 c^0 = 0, \quad \bm x \in \Omega_f, 
\] 
\[
 d^0 = 0, \quad \bm x \in \Gamma_s . 
\]  

\subsection{Linear kinetics of oxidation} 

First, we consider the oxidation process at the initial stage that is characterized by linear kinetics.
In this simplest case, predictions are performed for the following parameter set:
\[
 \mathrm{Re} = 50,
 \quad \mathrm{Pe} = 10,
 \quad \mathrm{Sh}_1 = 0.001,
 \quad \mathrm{Sh}_2^{-1} = 0.  
\]
Time--integration process is considered up to the moment $T = 50$ with the time--step $\tau =0.1$.
The average oxygen concentration at the outlet is treated as the integral characteristic of the process:
\begin{equation}\label{33}
 c_{out}(t) = \frac{\int_{\Gamma_{out}} c(\bm x, t) d \bm x}{\int_{\Gamma_{out}} d \bm x} . 
\end{equation} 

\begin{sidewaysfigure}   
  \begin{center}

    \includegraphics[width=1.1\linewidth] {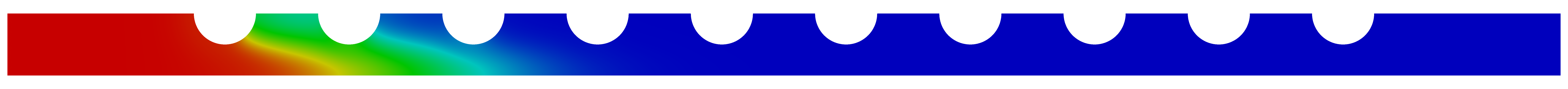}

	\vspace{5mm} a \vspace{5mm} 

    \includegraphics[width=1.1\linewidth] {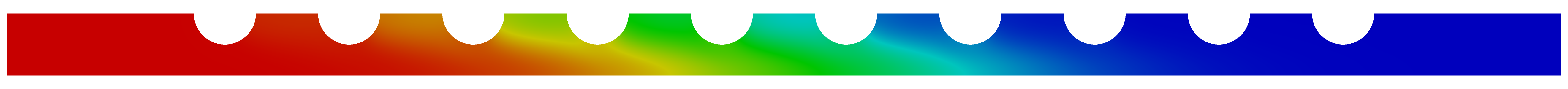}

	\vspace{5mm} b \vspace{5mm} 

    \includegraphics[width=1.1\linewidth] {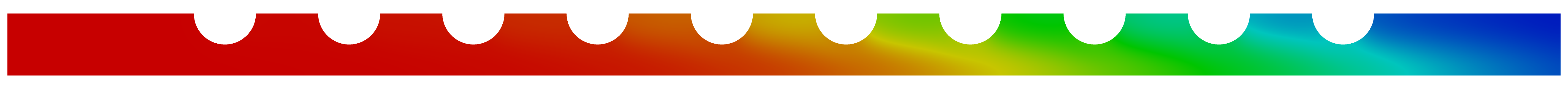}

	\vspace{5mm} c \vspace{5mm} 

	\caption{Distribution of the concentration $c$ for the in--line configuration at various time--moments: 
	         a --- $t = 5$,
                 b --- $t = 10$, 
                 c --- $t = 15$}
	\label{f-13}
  \end{center}
\end{sidewaysfigure}

\begin{sidewaysfigure}   
  \begin{center}

    \includegraphics[width=1.1\linewidth] {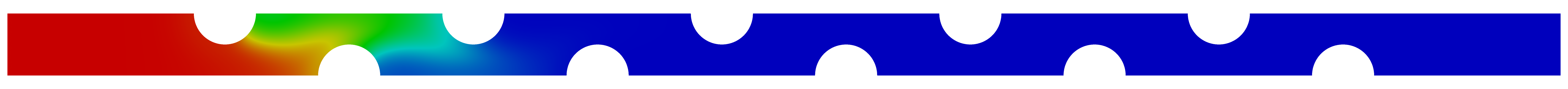}

	\vspace{5mm} a \vspace{5mm} 

    \includegraphics[width=1.1\linewidth] {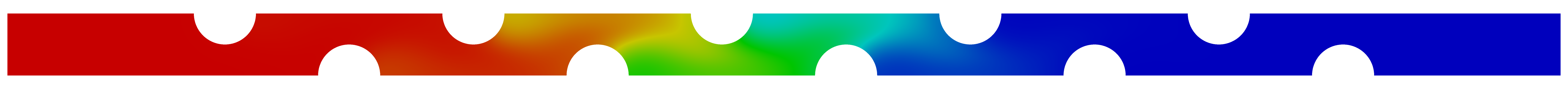}

	\vspace{5mm} b \vspace{5mm} 

    \includegraphics[width=1.1\linewidth] {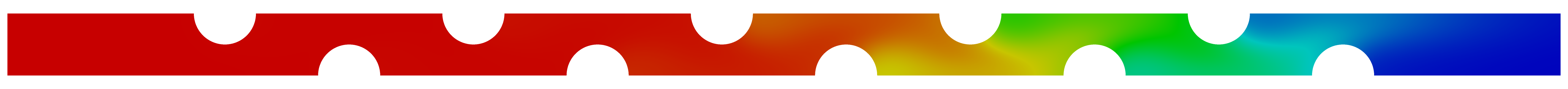}

	\vspace{5mm} c \vspace{5mm} 

	\caption{Distribution of the concentration $c$ for the staggered configuration at various time--moments: 
	         a --- $t = 5$,
                 b --- $t = 10$, 
                 c --- $t = 15$}
	\label{f-14}
  \end{center}
\end{sidewaysfigure}

\begin{sidewaysfigure}   
  \begin{center}

    \includegraphics[width=1.1\linewidth] {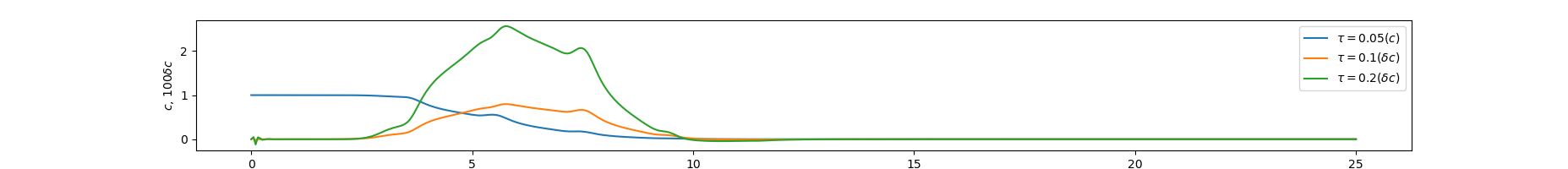}

	\vspace{5mm} a \vspace{5mm} 

    \includegraphics[width=1.1\linewidth] {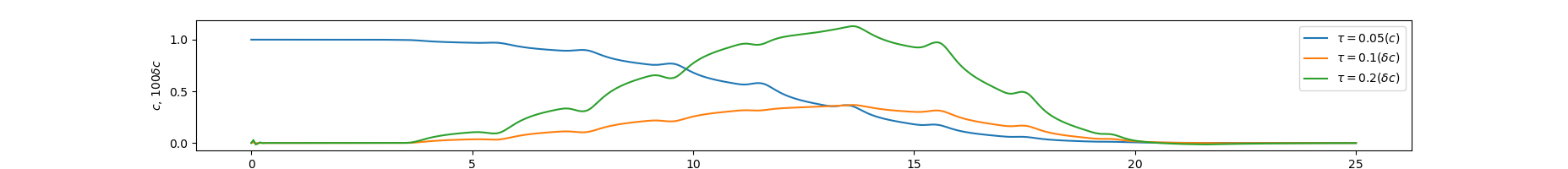}

	\vspace{5mm} b \vspace{5mm} 

    \includegraphics[width=1.1\linewidth] {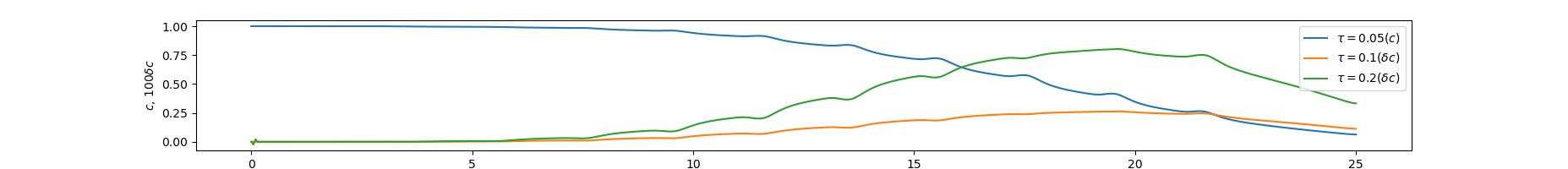}

	\vspace{5mm} c \vspace{5mm} 

	\caption{Concentration $c$ along the midline of the computational domain
	         for the in--line configuration presented as the solution obtained with $\tau = 0.05$ and 
	         the normalized deviations from it $\delta c$ obtained with $\tau = 0.1$ and $\tau = 0.2$
	         at various time--moments:
	         a --- $t = 5$,
                 b --- $t = 10$, 
                 c --- $t = 15$}
	\label{f-15}
  \end{center}
\end{sidewaysfigure}

\begin{sidewaysfigure}   
  \begin{center}

    \includegraphics[width=1.1\linewidth] {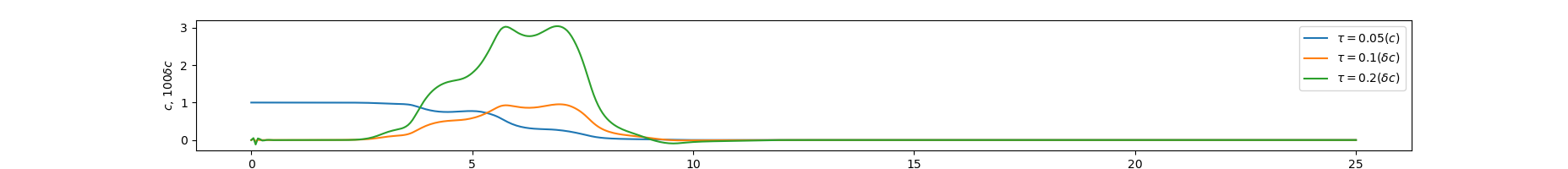}

	\vspace{5mm} a \vspace{5mm} 

    \includegraphics[width=1.1\linewidth] {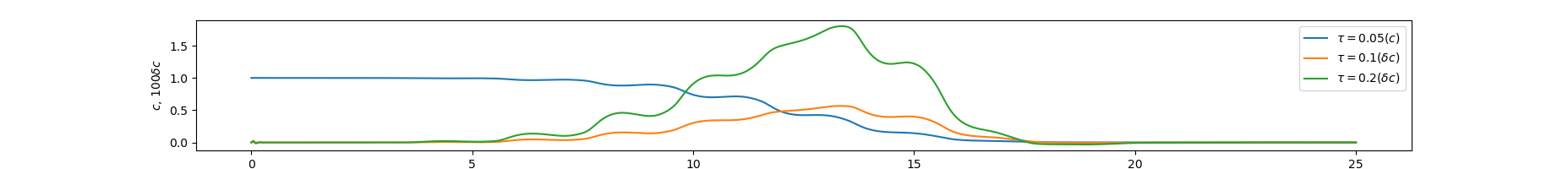}

	\vspace{5mm} b \vspace{5mm} 

    \includegraphics[width=1.1\linewidth] {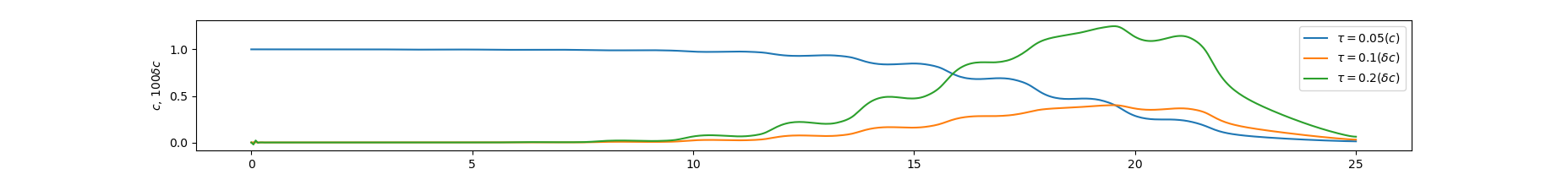}

	\vspace{5mm} c \vspace{5mm} 

        \caption{Concentration $c$ along the midline of the computational domain
	         for the staggered configuration presented as the solution obtained with $\tau = 0.05$ and 
	         the normalized deviations from it $\delta c$ obtained with $\tau = 0.1$ and $\tau = 0.2$
	         at various time--moments:
	         a --- $t = 5$,
                 b --- $t = 10$, 
                 c --- $t = 15$}
	\label{f-16}
  \end{center}
\end{sidewaysfigure}

The concentration distribution at different time--moments is shown in Figs.~\ref{f-13} and \ref{f-14}
for the in--line and staggered configuration, respectively. Figures~\ref{f-15} and \ref{f-16}
demonstrate the influence of the time--step on the solution for two configurations. In these figures,
the concentration $c$ is shown along the midline of the computational domain. 
These main plots are obtained on the finest grid in time corresponding to the time step $\tau = 0.05$. 
For a comparison, there is depicted the deviation $\delta c$ from this solution.
These two deviations obtained with $\tau = 0.1$ and $\tau = 0.2$ are normalized (multiplied by 100)
in order to make this comparison more evident. It is easy to see that the time--grid with $\tau =0.1$
provides a good enough accuracy for the time--integration. 

\begin{figure}[htp]
\begin{minipage}{0.49\linewidth}
  \begin{center}
    \includegraphics[width=1.\linewidth] {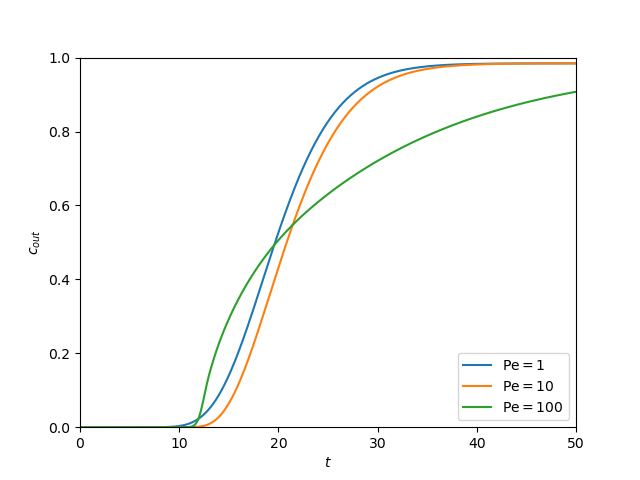}
  \end{center}
\end{minipage}\hfill
\begin{minipage}{0.49\linewidth}
  \begin{center}
    \includegraphics[width=1.\linewidth] {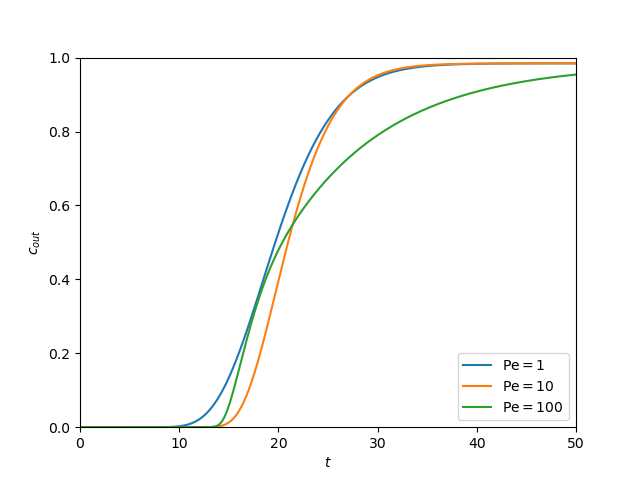}
  \end{center}
\end{minipage}
	\caption{Influence of $\mathrm{Pe}$: 
	 left --- in--line configuration, 
         right --- staggered configuration}
	\label{f-17}
\end{figure}

\begin{figure}[htp]
\begin{minipage}{0.49\linewidth}
  \begin{center}
    \includegraphics[width=1.\linewidth] {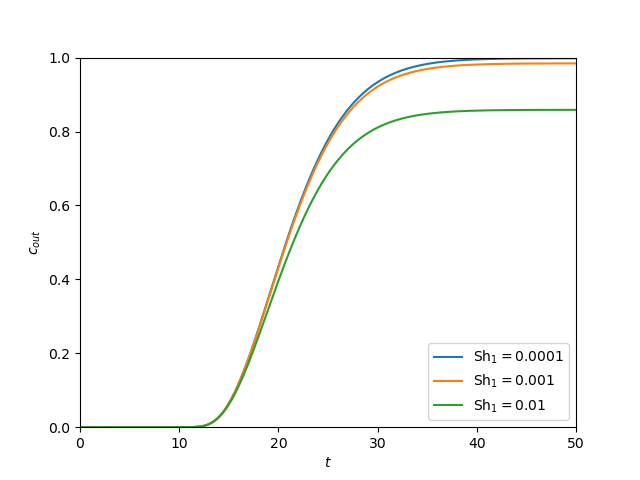}
  \end{center}
\end{minipage}\hfill
\begin{minipage}{0.49\linewidth}
  \begin{center}
    \includegraphics[width=1.\linewidth] {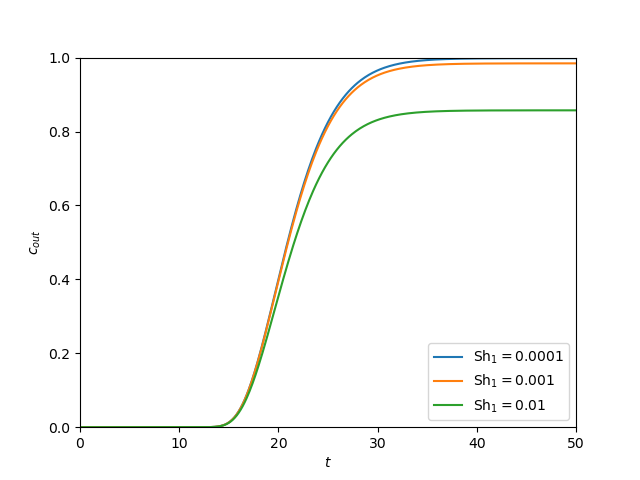}
  \end{center}
\end{minipage}
	\caption{Effect of $\mathrm{Sh}_1$: 
	 left --- in--line configuration,
         right --- staggered configuration}
	\label{f-18}
\end{figure}

The dependence of the average concentration at the outlet $c_{out}$ defined according to (\ref{33})
on the Peclet number is shown in Fig.~\ref{f-17}. The influence of the Sherwood number is given in Fig.~\ref{f-18}.

\begin{figure}[ht] 
  \begin{center}
    \begin{tikzpicture}[scale=0.75]
       \shade[top color=blue!20, bottom color=blue!0] (0,-1) rectangle +(10,4);
       \fill [fill=green!10,draw=black] (5,3) -- (2,3) arc (180:220:3) -- cycle;
       \fill [fill=green!20,draw=black] (5,3) -- (8,3) arc (360:220:3) -- cycle;
       \draw [-] (0,3) -- (10,3);
       \draw [->, line width=2, color=blue] (-0.5,0) -- (0.5,0);
       \draw [->, line width=2, color=blue] (-0.5,1) -- (0.5,1);
       \draw [->, line width=2, color=blue] (-0.5,2) -- (0.5,2);
       \draw[font=\LARGE]  (3.5,2.5) node {$\theta$};
    \end{tikzpicture}
    \caption{Local angle $\theta$} 
   \label{f-19}
  \end{center}
\end{figure}
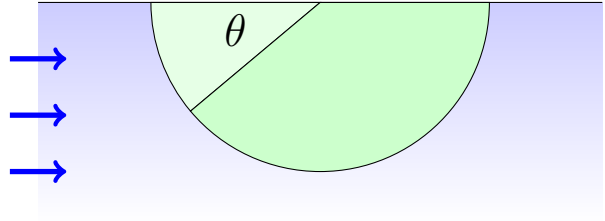

\begin{figure}[htp]
\begin{minipage}{0.49\linewidth}
  \begin{center}
    \includegraphics[width=1.\linewidth] {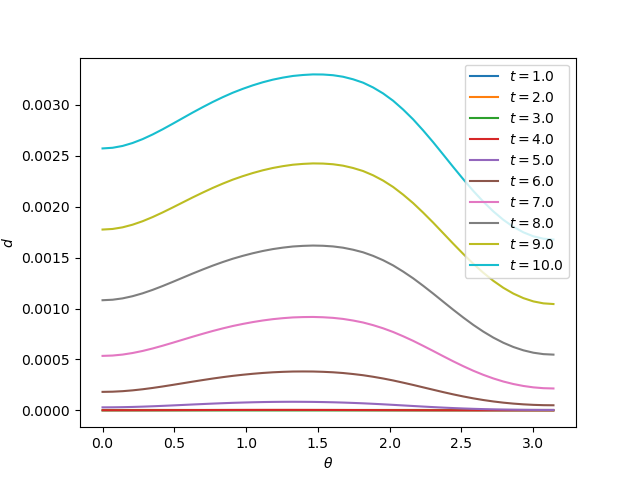}
  \end{center}
\end{minipage}\hfill
\begin{minipage}{0.49\linewidth}
  \begin{center}
    \includegraphics[width=1.\linewidth] {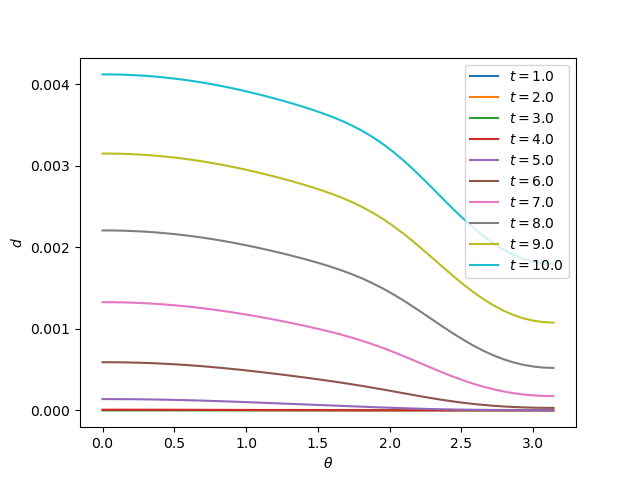}
  \end{center}
\end{minipage}
	\caption{Distribution of the oxide film thickness on the 3-rd tube surface: 
	 left --- in--line configuration, 
         right --- staggered configuration}
	\label{f-20}
\end{figure}

It is interesting to analyze the time--evolution of the local thickness of the oxide film on tubes.
The distribution of the film thickness along tube surfaces is presented in the dependence on 
the local angle $\theta$ defined in Fig.~\ref{f-19}. The value $\theta = 0$ corresponds to the leading edge of the tube,
whereas $\theta = \pi$ corresponds to the rear edge. The film thickness on the surface of the third (from the inlet)
tube is shown in Fig.~\ref{f-20} at various time--moments. We observe 
a higher growth rate of the film at the leading edge in comparison with the rear edge.
Next, for the staggered configuration, the maximum growth of the film takes place at the leading edge.
For the in--line configuration, the maximum growth is near the midpoint of the tube surface.
At high time-moments, the growth rate of the film tends to be constant.

\clearpage

\subsection{Effect of parabolic kinetics} 

In the general case, both linear and parabolic kinetics ($\mathrm{Sh}_2^{-1} > 0$) are taken into account.
Under these conditions, the problem at the new time level (see (\ref{31}), (\ref{32})) is nonlinear. 
To solve it, Newton's method is applied. As a rule, two or three iterations are necessary for the solution convergence
in calculations presented here.

\begin{figure}[htp]
\begin{minipage}{0.49\linewidth}
  \begin{center}
    \includegraphics[width=1.\linewidth] {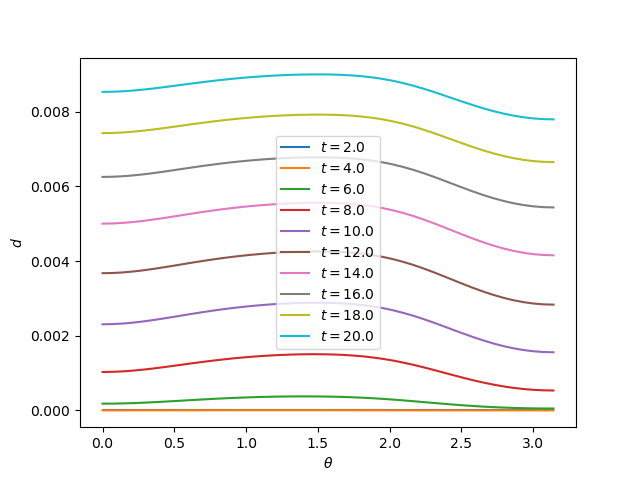}
  \end{center}
\end{minipage}\hfill
\begin{minipage}{0.49\linewidth}
  \begin{center}
    \includegraphics[width=1.\linewidth] {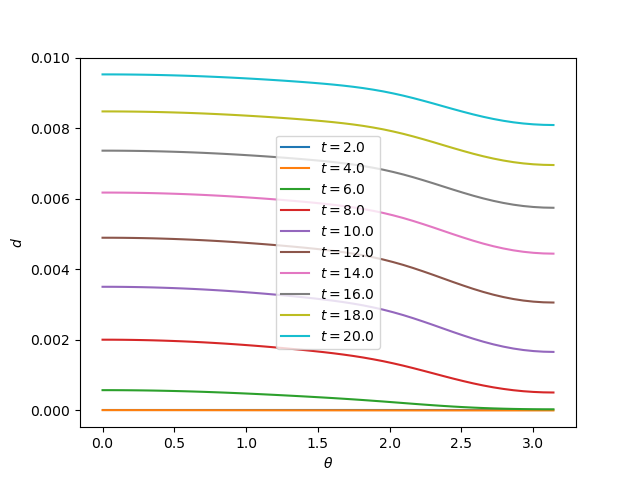}
  \end{center}
\end{minipage}
	\caption{Distribution of the oxide film thickness for $\mathrm{Sh}_2 = 10^{-5}$:
	        left --- in-line configuration, 
                right --- staggered configuration}
	\label{f-21}
\end{figure}

\begin{figure}[htp]
\begin{minipage}{0.49\linewidth}
  \begin{center}
    \includegraphics[width=1.\linewidth] {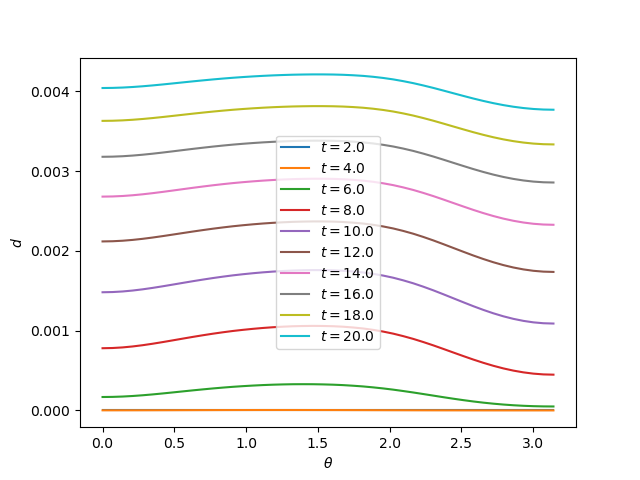}
  \end{center}
\end{minipage}\hfill
\begin{minipage}{0.49\linewidth}
  \begin{center}
    \includegraphics[width=1.\linewidth] {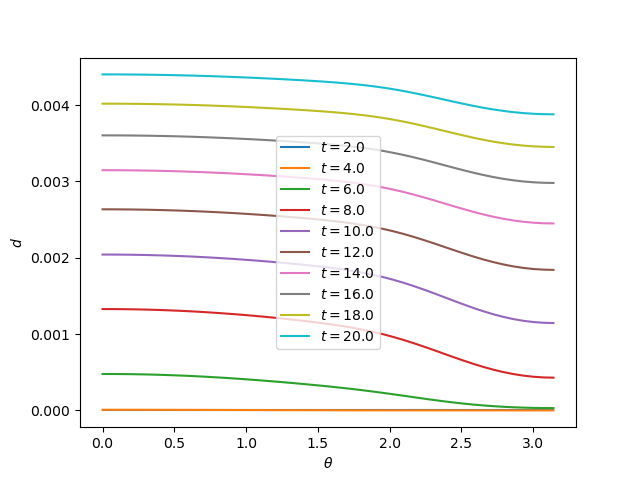}
  \end{center}
\end{minipage}
	\caption{Distribution of the oxide film thickness for $\mathrm{Sh}_2 = 10^{-6}$:            
	         left --- in-line configuration, 
                 right --- staggered configuration}
	\label{f-22}
\end{figure}

\begin{figure}[htp]
\begin{minipage}{0.49\linewidth}
  \begin{center}
    \includegraphics[width=1.\linewidth] {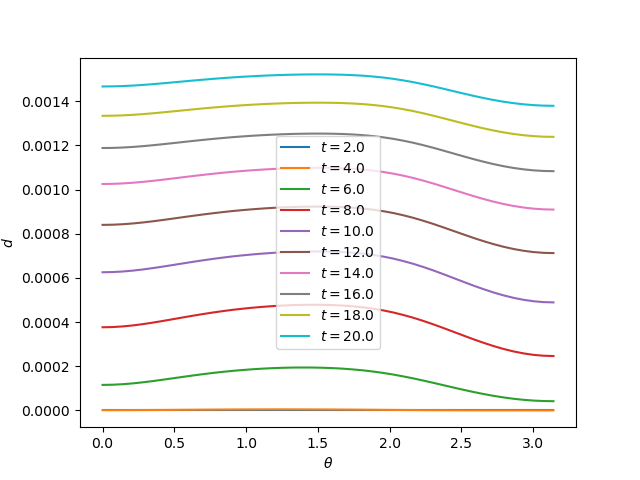}
  \end{center}
\end{minipage}\hfill
\begin{minipage}{0.49\linewidth}
  \begin{center}
    \includegraphics[width=1.\linewidth] {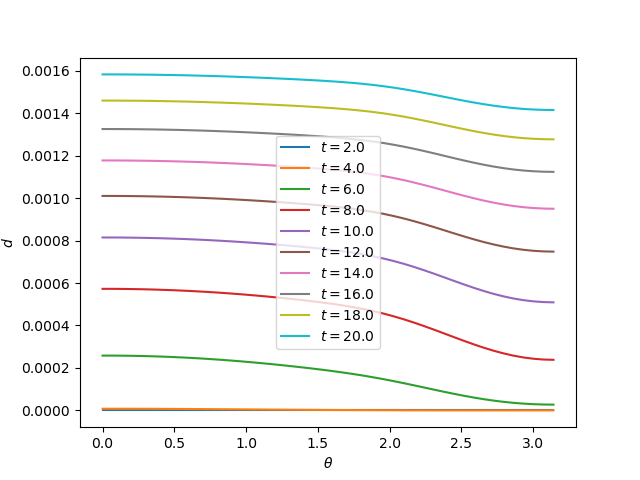}
  \end{center}
\end{minipage}
	\caption{Distribution of the oxide film thickness for $\mathrm{Sh}_2 = 10^{-7}$: 
	         left --- in-line configuration, 
                 right --- staggered configuration}
	\label{f-23}
\end{figure}

The impact of $\mathrm{Sh}_2$ is presented in Figs.~\ref{f-21}--\ref{f-23}.
These calculations are performed for $\mathrm{Sh}_1 = 0.001$.  
As $\mathrm{Sh}_2$ decreases, the growth rate of the oxide film also decreases.
This is due to the effect of the film thickness on the oxidation process.

\begin{figure}[htp]
\begin{minipage}{0.49\linewidth}
  \begin{center}
    \includegraphics[width=1.\linewidth] {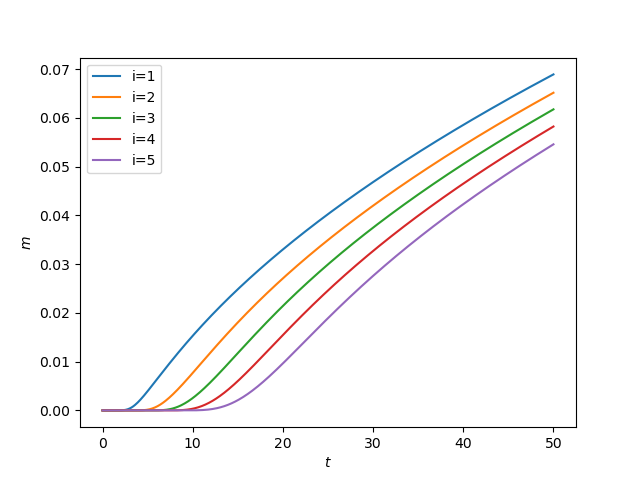}
  \end{center}
\end{minipage}\hfill
\begin{minipage}{0.49\linewidth}
  \begin{center}
    \includegraphics[width=1.\linewidth] {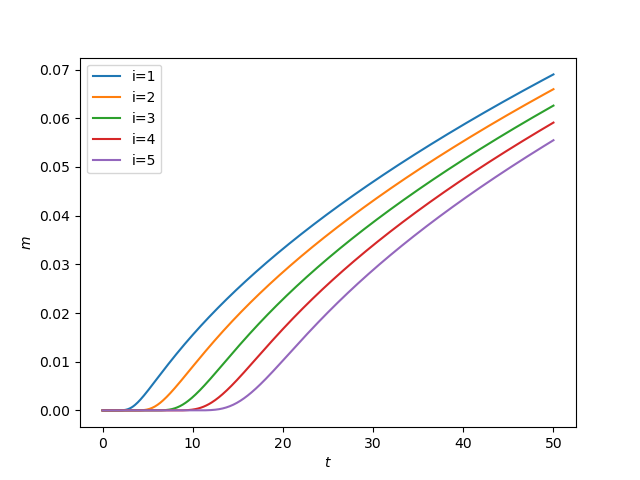}
  \end{center}
\end{minipage}
	\caption{Time--evolution of the oxide mass for the first five tubes at $\mathrm{Sh}_2 = 10^{-5}$: 
	         left --- in-line configuration, 
                 right --- staggered configuration}
	\label{f-24}
\end{figure}

\begin{figure}[htp]
\begin{minipage}{0.49\linewidth}
  \begin{center}
    \includegraphics[width=1.\linewidth] {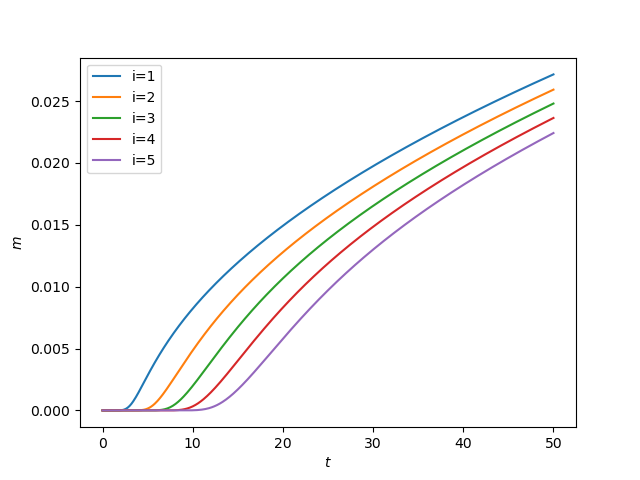}
  \end{center}
\end{minipage}\hfill
\begin{minipage}{0.49\linewidth}
  \begin{center}
    \includegraphics[width=1.\linewidth] {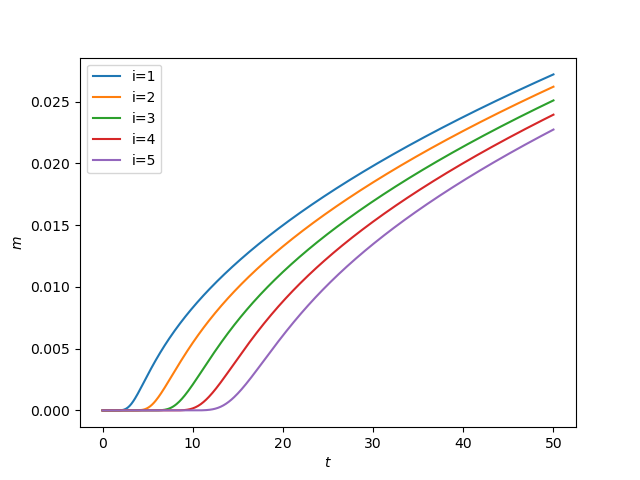}
  \end{center}
\end{minipage}
        \caption{Time--evolution of the oxide mass for the first five tubes at $\mathrm{Sh}_2 = 10^{-6}$: 
	         left --- in-line configuration, 
                 right --- staggered configuration}
	\label{f-25}
\end{figure}

\begin{figure}[htp]
\begin{minipage}{0.49\linewidth}
  \begin{center}
    \includegraphics[width=1.\linewidth] {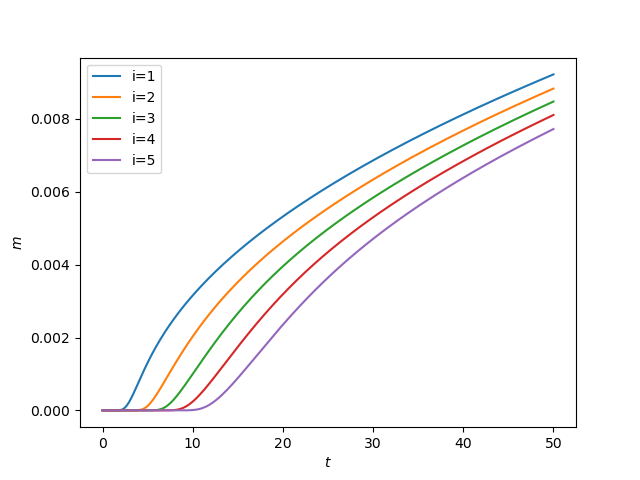}
  \end{center}
\end{minipage}\hfill
\begin{minipage}{0.49\linewidth}
  \begin{center}
    \includegraphics[width=1.\linewidth] {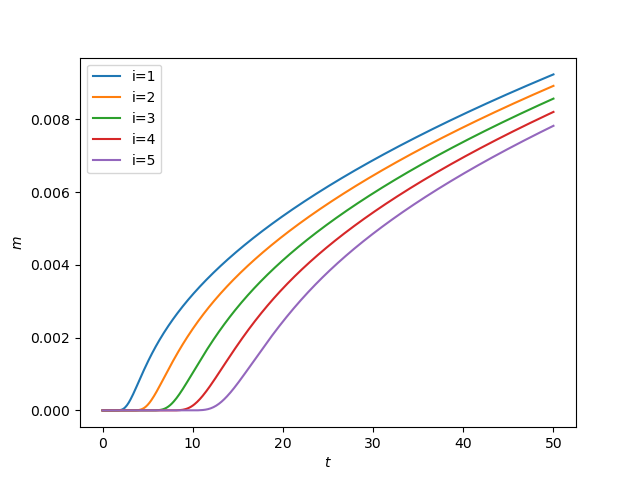}
  \end{center}
\end{minipage}
        \caption{Time--evolution of the oxide mass for the first five tubes at $\mathrm{Sh}_2 = 10^{-7}$: 
	         left --- in-line configuration,
	         right --- staggered configuration}
	\label{f-26}
\end{figure}

It is interesting to study the time--evolution of the oxide mass on an individual tube.
For the first five tubes in the in--line and staggered configurations with the boundaries $\Gamma_{s,i}$, we calculate
\[
 m_i(t) = 2 \int_{\Gamma_{s,i}} d (\bm x,t) d \bm x ,
 \quad i = 1,2, ..., 5 . 
\]
The dependence of the oxide mass on the number $\mathrm{Sh}_2$ is shown in Figs.~\ref{f-24}--\ref{f-26}.
There is no essential differences between the in--line and staggered configurations.
 
\clearpage

\section{Conclusions} 

A new mathematical model of the oxidation process is developed for a cross--flow around tube bundles.
It is based on the incompressible Navier--Stokes equations for laminar flows.
The oxidant transport is described by the convection--diffusion equation.
The peculiarity of the developed mass transfer model consists in the formulation of boundary conditions 
for the oxidant. They include explicitly the thickness of the oxide film and involve
both linear and parabolic kinetics of oxidation.

The computational algorithm is based on finite--element discretization in space using
Lagrangian finite elements on triangular grids.
Hydrodynamics is calculated independently of mass transfer in the stationary formulation. 
The Newton method is applied to solve governing equations in the primitive variables.
The Crank-Nicolson scheme is used for time--stepping in solving equations for the oxygen concentration 
and oxide film thickness.

Two--dimensional predictions of stationary laminar flows in in--line and staggered tube bundles are conducted
for different values of the Reynolds number. The grid independence study is performed using
a sequence of refined grids.

Calculations of the oxidation processes are carried out for the initial stage of oxidation 
in the linear kinetics approximation. A parametric study is done to indicate the influence of the Peclet and
Sherwood numbers. For the complete model that involves both linear and parabolic kinetics,
the mass transfer phenomenon is investigated numerically with emphasis on the time--evolution of the local 
thickness of the oxide film on tube surfaces.

\section*{Acknowledgements}

The work of the fourth author was supported by the grant of the Russian Federation Government (\#~14.Y26.31.0013).


\begin{thebibliography}{31}
\expandafter\ifx\csname natexlab\endcsname\relax\def\natexlab#1{#1}\fi
\providecommand{\url}[1]{\texttt{#1}}
\providecommand{\href}[2]{#2}
\providecommand{\path}[1]{#1}
\providecommand{\DOIprefix}{doi:}
\providecommand{\ArXivprefix}{arXiv:}
\providecommand{\URLprefix}{URL: }
\providecommand{\Pubmedprefix}{pmid:}
\providecommand{\doi}[1]{\href{http://dx.doi.org/#1}{\path{#1}}}
\providecommand{\Pubmed}[1]{\href{pmid:#1}{\path{#1}}}
\providecommand{\bibinfo}[2]{#2}
\ifx\xfnm\relax \def\xfnm[#1]{\unskip,\space#1}\fi
\bibitem[{Kays and London(1964)}]{kays1998compact}
\bibinfo{author}{W.~M. Kays}, \bibinfo{author}{A.~L. London},
  \bibinfo{title}{Compact Heat Exchangers}, \bibinfo{publisher}{McGraw-Hill},
  \bibinfo{address}{New York}, \bibinfo{year}{1964}.
\bibitem[{Bergman et~al.(2011)Bergman, Lavine, Incropera, and
  DeWitt}]{bergman2011fundamentals}
\bibinfo{author}{T.~L. Bergman}, \bibinfo{author}{A.~S. Lavine},
  \bibinfo{author}{F.~P. Incropera}, \bibinfo{author}{D.~P. DeWitt},
  \bibinfo{title}{Fundamentals of Heat and Mass Transfer},
  \bibinfo{publisher}{John Wiley \& Sons}, \bibinfo{year}{2011}.
\bibitem[{{\v{Z}}ukauskas(1972)}]{vzukauskas1972heat}
\bibinfo{author}{A.~{\v{Z}}ukauskas}, \bibinfo{journal}{Advances in Heat
  Transfer} \bibinfo{volume}{8} (\bibinfo{year}{1972})
  \bibinfo{pages}{93--160}.
\bibitem[{Iwaki et~al.(2004)Iwaki, Cheong, Monji, and Matsui}]{iwaki2004piv}
\bibinfo{author}{C.~Iwaki}, \bibinfo{author}{K.~H. Cheong},
  \bibinfo{author}{H.~Monji}, \bibinfo{author}{G.~Matsui},
  \bibinfo{journal}{Experiments in Fluids} \bibinfo{volume}{37}
  (\bibinfo{year}{2004}) \bibinfo{pages}{350--363}.
\bibitem[{Paul et~al.(2007)Paul, Tachie, and Ormiston}]{paul2007experimental}
\bibinfo{author}{S.~S. Paul}, \bibinfo{author}{M.~F. Tachie},
  \bibinfo{author}{S.~J. Ormiston}, \bibinfo{journal}{International Journal of
  Heat and Fluid Flow} \bibinfo{volume}{28} (\bibinfo{year}{2007})
  \bibinfo{pages}{441--453}.
\bibitem[{Beale and Spalding(1999)}]{beale1999numerical}
\bibinfo{author}{S.~B. Beale}, \bibinfo{author}{D.~B. Spalding},
  \bibinfo{journal}{Journal of Fluids and Structures} \bibinfo{volume}{13}
  (\bibinfo{year}{1999}) \bibinfo{pages}{723--754}.
\bibitem[{Wang et~al.(2000)Wang, Penner, and Ormiston}]{wang2000analysis}
\bibinfo{author}{Y.~Q. Wang}, \bibinfo{author}{L.~A. Penner},
  \bibinfo{author}{S.~J. Ormiston}, \bibinfo{journal}{Numerical Heat Transfer:
  Part A: Applications} \bibinfo{volume}{38} (\bibinfo{year}{2000})
  \bibinfo{pages}{819--845}.
\bibitem[{El-Shaboury and Ormiston(2005)}]{el2005analysis}
\bibinfo{author}{A.~M.~F. El-Shaboury}, \bibinfo{author}{S.~J. Ormiston},
  \bibinfo{journal}{Numerical Heat Transfer, Part A: Applications}
  \bibinfo{volume}{48} (\bibinfo{year}{2005}) \bibinfo{pages}{99--126}.
\bibitem[{Jayavel and Tiwari(2009)}]{jayavel2009numerical}
\bibinfo{author}{S.~Jayavel}, \bibinfo{author}{S.~Tiwari},
  \bibinfo{journal}{International Journal of Numerical Methods for Heat \&
  Fluid Flow} \bibinfo{volume}{19} (\bibinfo{year}{2009})
  \bibinfo{pages}{931--949}.
\bibitem[{Gowda et~al.(1998)Gowda, Patnaik, Narayana, and
  Seetharamu}]{gowda1998finite}
\bibinfo{author}{Y.~T.~K. Gowda}, \bibinfo{author}{B.~S. V.~P. Patnaik},
  \bibinfo{author}{P.~A.~A. Narayana}, \bibinfo{author}{K.~N. Seetharamu},
  \bibinfo{journal}{International Journal of Heat and Fluid Flow}
  \bibinfo{volume}{19} (\bibinfo{year}{1998}) \bibinfo{pages}{49--55}.
\bibitem[{Zdravistch et~al.(1995)Zdravistch, Fletcher, and
  Behnia}]{zdravistch1995numerical}
\bibinfo{author}{F.~Zdravistch}, \bibinfo{author}{C.~A. Fletcher},
  \bibinfo{author}{M.~Behnia}, \bibinfo{journal}{International Journal of
  Numerical Methods for Heat \& Fluid Flow} \bibinfo{volume}{5}
  (\bibinfo{year}{1995}) \bibinfo{pages}{717--733}.
\bibitem[{Dhaubhadel et~al.(1987)Dhaubhadel, Reddy, and
  Telionis}]{dhaubhadel1987finite}
\bibinfo{author}{M.~N. Dhaubhadel}, \bibinfo{author}{J.~N. Reddy},
  \bibinfo{author}{D.~P. Telionis}, \bibinfo{journal}{International Journal for
  Numerical Methods in Fluids} \bibinfo{volume}{7} (\bibinfo{year}{1987})
  \bibinfo{pages}{1325--1342}.
\bibitem[{Wilson and Bassiouny(2000)}]{wilson2000modeling}
\bibinfo{author}{A.~S. Wilson}, \bibinfo{author}{M.~K. Bassiouny},
  \bibinfo{journal}{Chemical Engineering and Processing: Process
  Intensification} \bibinfo{volume}{39} (\bibinfo{year}{2000})
  \bibinfo{pages}{1--14}.
\bibitem[{Khan et~al.(2006)Khan, Culham, and Yovanovich}]{khan2006convection}
\bibinfo{author}{W.~A. Khan}, \bibinfo{author}{J.~R. Culham},
  \bibinfo{author}{M.~M. Yovanovich}, \bibinfo{journal}{International Journal
  of Heat and Mass Transfer} \bibinfo{volume}{49} (\bibinfo{year}{2006})
  \bibinfo{pages}{4831--4838}.
\bibitem[{Lam et~al.(2010)Lam, Lin, Zou, and Liu}]{lam2010experimental}
\bibinfo{author}{K.~Lam}, \bibinfo{author}{Y.~F. Lin},
  \bibinfo{author}{L.~Zou}, \bibinfo{author}{Y.~Liu},
  \bibinfo{journal}{International Journal of Heat and Fluid Flow}
  \bibinfo{volume}{31} (\bibinfo{year}{2010}) \bibinfo{pages}{32--44}.
\bibitem[{Benarji et~al.(2008)Benarji, Balaji, and
  Venkateshan}]{benarji2008unsteady}
\bibinfo{author}{N.~Benarji}, \bibinfo{author}{C.~Balaji},
  \bibinfo{author}{S.~P. Venkateshan}, \bibinfo{journal}{Heat and Mass
  Transfer} \bibinfo{volume}{44} (\bibinfo{year}{2008})
  \bibinfo{pages}{445--461}.
\bibitem[{Horvat and Mavko(2006)}]{horvat2006heat}
\bibinfo{author}{A.~Horvat}, \bibinfo{author}{B.~Mavko},
  \bibinfo{journal}{Numerical Heat Transfer, Part A: Applications}
  \bibinfo{volume}{49} (\bibinfo{year}{2006}) \bibinfo{pages}{699--715}.
\bibitem[{Wang and Jackson(2010)}]{wang2010turbulence}
\bibinfo{author}{Y.~Q. Wang}, \bibinfo{author}{P.~L. Jackson},
  \bibinfo{journal}{Journal of Thermophysics and Heat Transfer}
  \bibinfo{volume}{24} (\bibinfo{year}{2010}) \bibinfo{pages}{534--543}.
\bibitem[{Tahseen et~al.(2015)Tahseen, Ishak, and Rahman}]{tahseen2015overview}
\bibinfo{author}{T.~A. Tahseen}, \bibinfo{author}{M.~Ishak},
  \bibinfo{author}{M.~M. Rahman}, \bibinfo{journal}{Renewable and Sustainable
  Energy Reviews} \bibinfo{volume}{43} (\bibinfo{year}{2015})
  \bibinfo{pages}{363--380}.
\bibitem[{Landau and Lifshitz(1987)}]{LandauLifshic1986}
\bibinfo{author}{L.~D. Landau}, \bibinfo{author}{E.~Lifshitz},
  \bibinfo{title}{Fluid Mechanics}, \bibinfo{publisher}{Pergamon Press},
  \bibinfo{address}{Oxford}, \bibinfo{year}{1987}.
\bibitem[{Li et~al.(2005)Li, Deen, and Kuipers}]{li2005numerical}
\bibinfo{author}{T.~Li}, \bibinfo{author}{N.~G. Deen},
  \bibinfo{author}{J.~A.~M. Kuipers}, \bibinfo{journal}{Chemical Engineering
  Science} \bibinfo{volume}{60} (\bibinfo{year}{2005})
  \bibinfo{pages}{1837--1847}.
\bibitem[{Young(2016)}]{young2016high}
\bibinfo{author}{D.~J. Young}, \bibinfo{title}{High Temperature Oxidation and
  Corrosion of Metals}, \bibinfo{publisher}{Elsevier},
  \bibinfo{address}{Amsterdam}, \bibinfo{year}{2016}.
\bibitem[{Deal and Grove(1965)}]{deal1965general}
\bibinfo{author}{B.~E. Deal}, \bibinfo{author}{A.~S. Grove},
  \bibinfo{journal}{Journal of Applied Physics} \bibinfo{volume}{36}
  (\bibinfo{year}{1965}) \bibinfo{pages}{3770--3778}.
\bibitem[{Xu et~al.(2011)Xu, Rosso, and Bruemmer}]{xu2011generalized}
\bibinfo{author}{Z.~Xu}, \bibinfo{author}{K.~M. Rosso}, \bibinfo{author}{S.~M.
  Bruemmer}, \bibinfo{journal}{The Journal of Chemical Physics}
  \bibinfo{volume}{135} (\bibinfo{year}{2011})
  \bibinfo{pages}{024108--1--024108--7}.
\bibitem[{Geuzaine and Remacle(2009)}]{Gmsh}
\bibinfo{author}{C.~Geuzaine}, \bibinfo{author}{J.-F. Remacle},
  \bibinfo{journal}{International Journal for Numerical Methods in Engineering}
  \bibinfo{volume}{79} (\bibinfo{year}{2009}) \bibinfo{pages}{1309--1331}.
  \DOIprefix\doi{10.1002/nme.2579}.
\bibitem[{Gresho and Sani(2000)}]{gresho200incompressible}
\bibinfo{author}{P.~M. Gresho}, \bibinfo{author}{R.~L. Sani},
  \bibinfo{title}{Incompressible Flow and the Finite Element Method, Volume 2,
  Isothermal Laminar Flow}, \bibinfo{publisher}{Wiley},
  \bibinfo{address}{Chichester}, \bibinfo{year}{2000}.
\bibitem[{Taylor and Hood(1973)}]{taylor1973numerical}
\bibinfo{author}{C.~Taylor}, \bibinfo{author}{P.~Hood},
  \bibinfo{journal}{Computers \& Fluids} \bibinfo{volume}{1}
  (\bibinfo{year}{1973}) \bibinfo{pages}{73--100}.
  \DOIprefix\doi{10.1016/0045-7930(73)90027-3}.
\bibitem[{Logg et~al.(2012)Logg, Mardal, Wells, and
  (Eds.)}]{LoggMardalEtAl2012a}
\bibinfo{author}{A.~Logg}, \bibinfo{author}{K.-A. Mardal},
  \bibinfo{author}{G.~N. Wells}, \bibinfo{author}{(Eds.)},
  \bibinfo{title}{Automated Solution of Differential Equations by the Finite
  Element Method: The FEniCS Book}, \bibinfo{publisher}{Springer-Verlag},
  \bibinfo{address}{Berlin}, \bibinfo{year}{2012}.
  \DOIprefix\doi{10.1007/978-3-642-23099-8}.
\bibitem[{Aln{\ae}s et~al.(2015)Aln{\ae}s, Blechta, Hake, Johansson, Kehlet,
  Logg, Richardson, Ring, Rognes, and Wells}]{AlnaesBlechta2015a}
\bibinfo{author}{M.~S. Aln{\ae}s}, \bibinfo{author}{J.~Blechta},
  \bibinfo{author}{J.~Hake}, \bibinfo{author}{A.~Johansson},
  \bibinfo{author}{B.~Kehlet}, \bibinfo{author}{A.~Logg},
  \bibinfo{author}{C.~Richardson}, \bibinfo{author}{J.~Ring},
  \bibinfo{author}{M.~E. Rognes}, \bibinfo{author}{G.~N. Wells},
  \bibinfo{journal}{Archive of Numerical Software} \bibinfo{volume}{3}
  (\bibinfo{year}{2015}) \bibinfo{pages}{9--23}.
  \DOIprefix\doi{10.11588/ans.2015.100.20553}.
\bibitem[{Samarskii(2001)}]{Samarskii}
\bibinfo{author}{A.~A. Samarskii}, \bibinfo{title}{The Theory of Difference
  Schemes}, \bibinfo{publisher}{Marcel Dekker, Inc.}, \bibinfo{address}{New
  York}, \bibinfo{year}{2001}.
\bibitem[{Ascher(2008)}]{Ascher2008}
\bibinfo{author}{U.~M. Ascher}, \bibinfo{title}{Numerical Methods for
  Evolutionary Differential Equations}, \bibinfo{publisher}{SIAM},
  \bibinfo{address}{Philadelphia, PA}, \bibinfo{year}{2008}.

\end{thebibliography}
\end{document}